\documentclass[lettersize,journal]{IEEEtran}
\usepackage{amsmath,amsfonts}
\usepackage{amsthm}
\usepackage{multirow}
\usepackage{adjustbox}
\usepackage{booktabs}
\usepackage{makecell}
\usepackage{algorithmic}

\usepackage{array}
\usepackage[table]{xcolor}
\usepackage[caption=false,font=normalsize]{subfig}
\usepackage{textcomp}
\usepackage[utf8]{inputenc}
\usepackage{amssymb}
\usepackage{stfloats}
\usepackage{url}
\usepackage{verbatim}
\usepackage{graphicx}
\usepackage{cite}
\usepackage{stackengine}
\usepackage{amsmath}
\usepackage{bbm}
\usepackage{mathtools}
\usepackage[ruled,vlined]{algorithm2e}
\usepackage{booktabs}
\usepackage{xcolor}
\usepackage{titlesec}
\newif\ifshowrevisions
\showrevisionsfalse
\long\def\revised#1{\ifshowrevisions{\color{blue}#1}\else#1\fi}

\usepackage{geometry}
\usepackage[hidelinks]{hyperref}

\def\delequal{\mathrel{\ensurestackMath{\stackon[1pt]{=}{\scriptstyle\Delta}}}}

\newtheorem{Proposition}{Proposition}

\begin{document}

\title{MORIC: CSI Delay-Doppler Decomposition for Robust Wi-Fi-based Human Activity Recognition}

\author{Navid Hasanzadeh$^{1}$ and Shahrokh Valaee$^{1}$,~\IEEEmembership{Fellow,~IEEE}

\thanks{$^{1}$N. Hasanzadeh and S. Valaee are with the Department of Electrical \& Computer Engineering, University of Toronto, Toronto, ON, Canada. \revised{navid.hasanzadeh@mail.utoronto.ca, valaee@ece.utoronto.ca.}}}



\maketitle

\begin{abstract}
The newly established \revised{IEEE 802.11bf} Task Group aims to amend the WLAN standard to support advanced sensing applications such as human activity recognition (HAR). Although studies have demonstrated the potential of sub-$7$ GHz Wi-Fi Channel State Information (CSI) for HAR, existing methods often degrade substantially under realistic variations across users, environments, and sensing configurations.
\revised{This work addresses the poor generalization of Wi-Fi-based HAR by extracting motion-centered representations that reduce dependence on static, environment-specific, and non-activity-related CSI magnitude and phase patterns. CSI signals are transformed into the delay-profile space and decomposed into multiple Doppler velocity projections, which are modeled as observations of a moving point’s velocity from different unknown directions, analogous to virtual cameras observing the same motion with varying degrees of clarity. This yields a richer activity representation than either a single aggregated Doppler estimate or the spurious, environment-dependent CSI patterns used in prior works. Since these projections are unordered and may recur due to random multipath propagation, we introduce MORIC, a novel order- and repetition-invariant time-series classification model for robust Wi-Fi-based HAR.
}
\revised{Experimental results on the collected dataset show that the proposed method outperforms state-of-the-art approaches in cross-user hand motion recognition, especially for challenging gestures. Incorporating only a few calibration samples further improves accuracy, demonstrating MORIC's adaptability and highlighting the potential of the proposed methodology for practical Wi-Fi sensing in real-world scenarios.}
\end{abstract}


\section{Introduction}
\IEEEPARstart{H}{uman} sensing and interaction technologies are transforming how humans interact with their environments, enabling devices to detect, interpret, and respond to their actions seamlessly. Among these technologies, Wi-Fi-based human activity recognition (HAR) is particularly promising due to its non-intrusive nature and widespread availability~\cite{ahmad2024wifi}. 
Using Wi-Fi signals, HAR systems can monitor human movements without requiring users to wear devices or interact directly with sensors.

The sensing capabilities of Wi-Fi have been explored across a broad spectrum of applications, ranging from smart homes to healthcare, encompassing areas such as gesture recognition~\cite{xu2025evaluating}, human tracking~\cite{shi2025ml}, people counting~\cite{jiang2023pa}, and sleep monitoring~\cite{alzaabi2025wi}. Despite its potential, Wi-Fi sensing remains significantly constrained by limited accuracy and poor generalization, making real-world deployment challenging. Furthermore, the current IEEE $802$ standard lacks dedicated sensing features, forcing Wi-Fi-based sensing solutions to rely on proprietary implementations that suffer from restricted interoperability~\cite{chen2022wi}.

To address these limitations, the newly introduced IEEE $802.11$bf amendment establishes standardized WLAN sensing procedures and enables sensing in license-exempt frequency bands below $7$ GHz~\cite{du2024overview}. This advancement is expected to bridge the gap between research and real-world adoption, paving the way for the seamless integration of sensing technologies into everyday Wi-Fi devices. In the near future, commercial HAR solutions could become a native feature of widely used Wi-Fi-enabled devices in homes, hospitals, schools, and workplaces~\cite{yousefi2017survey}. This transformation has the potential to revolutionize human sensing and interaction, with far-reaching applications in healthcare monitoring, smart environments, unlocking new possibilities for immersive digital experiences, and intelligent automation. 

HAR systems based on Wi-Fi exploit the impact of human movements on Channel State Information (CSI) between two or more Wi-Fi devices, capturing how wireless signals interact with static and dynamic elements in the environment~\cite{salehinejad2022litehar,ding2024multiple}. Previous studies have proposed a variety of Wi-Fi CSI-based methods for HAR~\cite{yousefi2017survey,djogoHAR,jang2025study,zhang2024csi,peng2023rosefi}, typically relying on either the magnitude or phase of CSI signals, combined with feature extraction and machine learning techniques. While magnitude-based methods achieve high accuracy in controlled settings, they are highly sensitive to environmental changes, limiting their robustness and practicality. Phase-based methods~\cite{wu2022wifi,yin2022fewsense} also suffer from phase wrapping, hardware distortions, and multipath interference. \revised{Despite efforts such as phase sanitization and ratio-based models, recent work~\cite{varga2024exposing} shows that both magnitude- and phase-based approaches degrade significantly under changes in users, environments, or devices, highlighting the cross-user and cross-environment generalization challenges that determine the practicality of Wi-Fi-based HAR for real-world deployment.}

\revised{An alternative line of work seeks to extract more physically meaningful representations of human activities and motion from Wi-Fi CSI, such as motion velocity. These methods exploit the frequency shift induced by human movement in Wi-Fi signals, commonly referred to as Doppler shift or Doppler velocity. They estimate motion velocity by formulating the problem as an angle-of-arrival (AoA) estimation task~\cite{meneghello2022sharp, chen2022afall, zhang2021widar3}, where body motion is modeled as a single point source to recover directional velocity information. However, these approaches typically assume only one velocity measurement per access point (AP), thereby aggregating the contributions of multiple multipath components into a single Doppler representation. Such a single Doppler representation may be insufficient to distinguish between activities with similar projected motion patterns, especially when different gestures produce overlapping velocity signatures. This aggregation can further introduce ambiguity when distinguishing challenging human activities and gestures, since the dominant path, often the line-of-sight (LOS) component, may primarily reflect propagation conditions and material reflectivity rather than the full structure of the underlying motion. Moreover, multipath interference can suppress or distort essential motion cues. Although some studies incorporate multiple APs~\cite{zhang2021widar3}, each AP still produces a single velocity estimate, highlighting the need for techniques that isolate and interpret multipath-specific Doppler contributions to provide a more comprehensive representation of the performed activities and improve HAR accuracy.
}

\revised{Building on these limitations, this paper introduces \textbf{M}ultipath \textbf{O}rder and \textbf{R}epetition \textbf{I}nvariant \textbf{C}lassification (MORIC), a Wi-Fi-based HAR framework that aims to improve robustness and cross-user generalization by using Doppler velocity information from delay-separated multipath components rather than CSI magnitude, CSI phase, or a single aggregated Doppler estimate.}

\revised{The intuition behind MORIC is that generalizable HAR features should describe how the user moves, rather than the exact static channel through which the movement is observed. In indoor environments, Wi-Fi signals are reflected by walls, furniture, and other objects in a random manner. As a result, the same motion may be observed through different propagation paths, while similar path responses may appear multiple times or at different delay positions. MORIC is therefore designed to aggregate multiple motion-sensitive observations without assuming a fixed order or a fixed number of useful paths. Moreover, MORIC addresses the limited size of HAR datasets by using training-free temporal feature extraction before classification, reducing the need to train complex data-hungry deep models that are more prone to overfitting and thereby improving generalization.
}

\begin{figure}[!t]
	\centering
	\subfloat{%
		\includegraphics[width=0.7\linewidth]{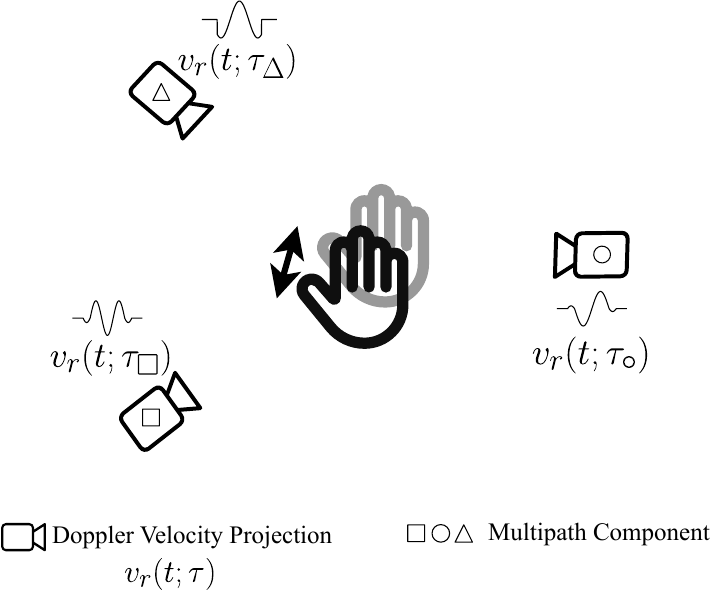}
	}
	\caption{MORIC represents each activity using the set of multipath Doppler velocity projections extracted from Wi-Fi CSI, providing a comprehensive motion descriptor that captures the activity from multiple perspectives within the $3$-D environment.
}
\label{figure:handcamera}
\end{figure}

The specific contributions of this work are summarized as follows:

\begin{itemize}
    \item This paper presents a novel method for extracting activity-related information by capturing Doppler velocity from multiple perspectives through delay-Doppler decomposition. Leveraging Wi-Fi's multipath propagation, each delay bin functions as a distinct one-dimensional virtual camera randomly positioned around the user (see Fig.~\ref{figure:handcamera}). By decomposing CSI into delay-separated multipath components, the method isolates motion-induced Doppler shifts, yielding a rich representation of the performed activity.

    \item Due to the inherent randomness of multipath Wi-Fi signal propagation, this paper proposes MORIC, a novel classifier that uses random convolutional kernels for time-series feature extraction from Doppler velocity projections and is designed to be invariant to the unpredictable ordering and repetition of multipath components, thereby ensuring robust outputs. By effectively addressing the stochastic nature of multipath propagation, MORIC significantly enhances generalization across different users, paving the way for Wi-Fi-based HAR in real-world scenarios.

    \item This paper evaluates the proposed method both theoretically and experimentally by modeling the effect of an arbitrarily moving point on CSI and validating performance on a challenging Wi-Fi-based hand motion dataset collected for this study. The results show that the method markedly outperforms existing approaches in generalization accuracy. An optional calibration procedure using only a few samples is also introduced to further mitigate user-specific variability.

\end{itemize}

\section{Background}\label{section:background}
\subsection{Wi-Fi Channel State Information}

CSI characterizes the impact of a wireless channel on transmitted signals.
Mathematically, for a transmitter with $A_t$ antennas and a receiver with $A_r$ antennas, the received signal vector $\mathbf{x} \in \mathbb{C}^{A_r}$ is related to the transmitted signal vector $\mathbf{s} \in \mathbb{C}^{A_t}$ and the channel matrix $\mathbf{H}_{f_c} \in \mathbb{C}^{A_r \times A_t}$ at subcarrier frequency $f_c$ by the equation
\begin{equation}
\mathbf{x} = \mathbf{H}_{f_c} \mathbf{s} + \mathbf{n},
\label{equation:9}
\end{equation}
where $\mathbf{n} \in \mathbb{C}^{A_r}$ is the additive noise. CSI estimation is typically performed using pilot symbols embedded in predefined training sequences known as \textit{Long Training Fields} (LTFs), which are included in the preamble of Wi-Fi packets. These LTFs span all subcarriers and are known at the receiver, allowing estimation of the channel. Let the pilot matrix be $\mathbf{P} \in \mathbb{C}^{A_t \times N_p}$, containing $N_p$ known pilot vectors $\mathbf{p_i}$. The corresponding received signals $\mathbf{x_i}$ are collected into matrix $\mathbf{X} \in \mathbb{C}^{A_r \times N_p}$, giving
\begin{equation}
\mathbf{X} = \mathbf{H}_{f_c} \mathbf{P} + \mathbf{N},
\end{equation}
with noise matrix $\mathbf{N} \in \mathbb{C}^{A_r \times N_p}$. The Least Squares (LS)~\cite{bjorck1990least} estimate of the CSI matrix is given by
\begin{equation}
\mathbf{H}_{f_c}^{\text{LS-estimate}} = \mathbf{X P^H (P P^H)^{-1}},
\end{equation}
where $(\cdot)^H$ denotes the Hermitian transpose. The estimation error is minimized when $\mathbf{P P^H}$ is proportional to the identity matrix, which is true if $N_p \geq A_t$ \cite{tulino2005impact}. Environmental changes modify the multipath structure, leading to time-varying estimates of $\mathbf{H}_{f_c}^{\text{LS-estimate}}$, which enables HAR by analyzing CSI sequences over time.

\subsection{Multipath Signal Propagation in CSI}
Multipath propagation is a fundamental aspect of wireless communication, where transmitted signals reach the receiver via multiple distinct paths due to reflections, scattering, and diffractions from objects like walls, furniture, or humans. Instead of following a single line-of-sight path, each signal copy travels a different route, causing variations in amplitude, delay, and phase. The wireless channel is thus modeled as the sum of these multipath components between the transmitter and receiver, as follows

\begin{equation}
\label{eq:CSI_multipath}
H_{m,n}^{f_c}(s) = \sum_{l=1}^L \beta_{l}(s)\,e^{-j2\pi d_{m,n,l}(s)\,f_c /c}
\end{equation}

\noindent
where \(H_{m,n}^{f_c}(s)\) denotes the CSI between the transmit antenna \(m\) and the receive antenna \(n\) at time \(s\). The summation runs over \(L\) distinct propagation paths. The term \(\beta_{l}(s) \in \mathbb{C}\) represents the complex gain associated with the \(l^{th}\) path, which captures its attenuation and phase shift due to interactions with the environment. The distance traveled by the \(l^{th}\) path is denoted by \(d_{m,n,l}(s)\), and the exponential term accounts for the phase shift caused by this propagation distance, where \(f_c\) is the carrier frequency and \(c\) is the speed of light. This model captures the aggregate effect of all propagation paths, highlighting how multipath dynamics contributes to the time-varying nature of Wi-Fi signal reception.

\section{Method} \label{section:method}

\revised{This section first describes Wi-Fi CSI noise elimination as an essential step for revealing activity-relevant patterns in CSI measurements for HAR. It then presents the main contributions of this paper by formulating the impact of human motion on CSI and introducing delay-separated Doppler representations as a comprehensive characterization of the performed activity. Based on this formulation, the section presents MORIC, a novel human activity and gesture classifier designed to handle the randomness introduced by indoor multipath propagation.}

\subsection{CSI Noise Elimination}
In practical Wi-Fi systems, various noise sources arise from different processing operations on the transmitter and receiver sides, as well as from hardware and software imperfections~\cite{ma2019wifi}. Taking into account these noise contributions, the measured CSI, $\hat{H}_{m,n}^{f_c}(s)$, at time \(s\) between an arbitrary transmit antenna \(m\) and the receive antenna \(n\) can be expressed as
\begin{equation}
\begin{multlined}
\hat{H}_{m,n}^{f_c}(s) = \underbrace{\sum_{l=1}^L \beta_{l}(s) \, e^{-j2\pi d_{m,n,l}(s) f_c/c}}_{\substack{\text{Multipath} \\ \text{Channel}}} \, \underbrace{e^{-j2\pi \tau_m(s) f_c}}_{\substack{\text{Cyclic Shift Diversity}}} \times \\
\underbrace{e^{-j2\pi \rho(s) f_c}}_{\substack{\text{Sampling Time Offset}}} \, \underbrace{e^{-j2\pi \eta(s) \left(\frac{f_c^\prime}{f_c} - 1\right) f_c}}_{\substack{\text{Sampling Frequency Offset}}} \, \underbrace{q_{m,n}(s) \, e^{-j2\pi \zeta_{m,n}(s)}}_{\text{Beamforming}},
\label{eq:CSI_channel}
\end{multlined}
\end{equation}
where the term \(\tau_m(s)\) denotes the delay introduced by cyclic shift diversity (CSD) at the \(m^\text{th}\) transmit antenna, \(\rho(s)\) represents the sampling time offset (STO), and \(\eta(s)\) is the sampling frequency offset (SFO) with \(f_c^\prime\) being the actual subcarrier frequency. Finally, \(q_{m,n}(s)\) and \(\zeta_{m,n}(s)\) correspond to the amplitude attenuation and phase shift introduced by the beamforming process, respectively.

\begin{figure}[!t]
	\centering
	\subfloat{%
		\includegraphics[width=0.7\linewidth]{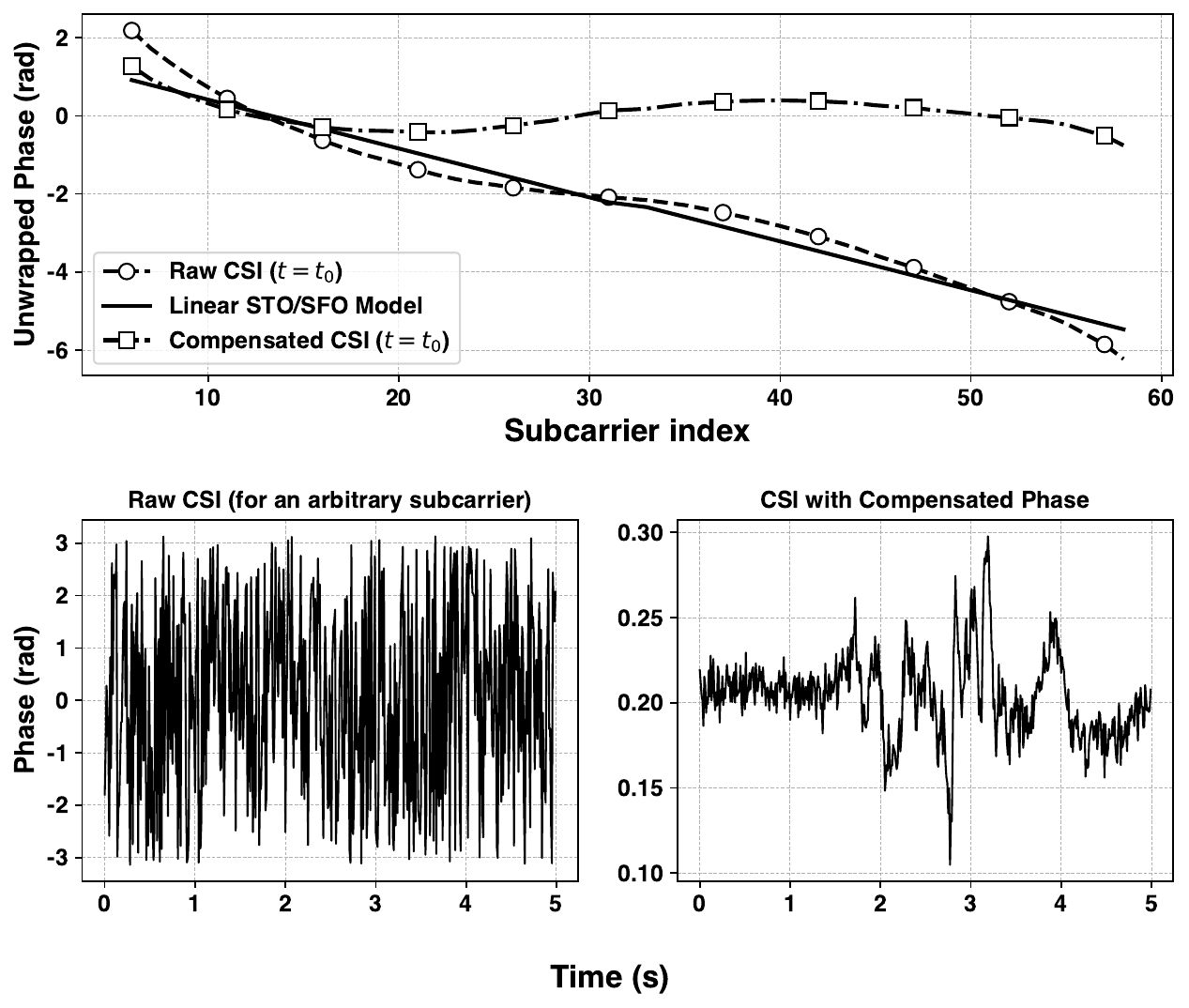}
	}
	\caption{\revised{Illustration of how phase sanitization reveals activity-related patterns in the CSI phase. 
Top: unwrapped raw CSI phase at time $t_0$, the fitted linear STO/SFO model, and the compensated CSI phase across subcarriers. 
The preprocessing step is performed by subtracting the fitted linear STO/SFO component from the raw CSI phase. 
Bottom left: raw CSI phase before preprocessing, which is noisy and does not show an identifiable activity pattern. 
Bottom right: compensated CSI phase over time, showing activity-related patterns.}}

\label{figure:noise}
\end{figure}

STO and SFO are known to be especially destructive to the CSI phase compared to other hardware or software imperfections. The phase information, which contains valuable Doppler-induced details, is heavily corrupted by these offsets, while other noise sources typically manifest as minor temporal variations. Unlike some previous CSI processing methods that rely on averaging or low-pass filtering to eliminate temporal noise, such techniques may exacerbate the impact of unpredictable phase spikes and jumps caused by synchronization errors, hardware limitations, or environmental disturbances.

To mitigate the impact of STO and SFO, an effective phase compensation method is proposed in \cite{tadayon2019decimeter} as a pre-processing step for CSI measurements. At a given time instant \(s\), let \(\hat{\theta}_{s,k}\) denote the measured phase in the \(k^{th}\) subcarrier. A linear regression model is applied along the subcarrier dimension to estimate the combined effects of STO and SFO. The linear model is given by
\begin{equation}
r_s(k) = \epsilon(s) \cdot k + \tau(s),
\label{eq:linreg_t}
\end{equation}
where the slope \(\epsilon(s)\) and the offset \(\tau(s)\) are estimated as
\begin{equation}
\epsilon(s) = \frac{\sum_{k=1}^K \left(\hat{\theta}_{s,k} - \bar{\hat{\Theta}}_{s,*}\right)(k-\bar{k})}{\sum_{k=1}^K (k-\bar{k})^2},
\label{eq:slope_t}
\end{equation}
\begin{equation}
\tau(s) = \bar{\hat{\Theta}}_{s,*} - \bar{k} \cdot \epsilon(s),
\label{eq:offset_t}
\end{equation}
with \(\bar{\hat{\Theta}}_{s,*}\) denoting the average measured phase across all subcarriers at time \(s\) and \(\bar{k}\) the average subcarrier index. The sanitized phase is then obtained by subtracting the estimated linear trend
\begin{equation}
\theta'_s(k) = \hat{\theta}_{s,k} - \epsilon(s)\, k - \tau(s).
\label{eq:corrected_phase_t}
\end{equation}
As shown in Fig.~\ref{figure:noise}, the phase-sanitization step in Eq.~\eqref{eq:corrected_phase_t} subtracts the fitted linear STO/SFO component from the unwrapped raw CSI phase. This compensation effectively removes the dominant linear phase distortions across subcarriers while preserving the underlying multipath channel structure. As a result, activity-related temporal patterns that are obscured in the raw phase become more visible in the compensated CSI phase, providing useful motion information for accurate HAR.

\subsection{The Effect of Motion on CSI}
\begin{figure}[!t]
    \centering
    \subfloat[$T\!\rightarrow\!R$\label{fig:geo:a}]{%
        \includegraphics[width=0.48\linewidth]{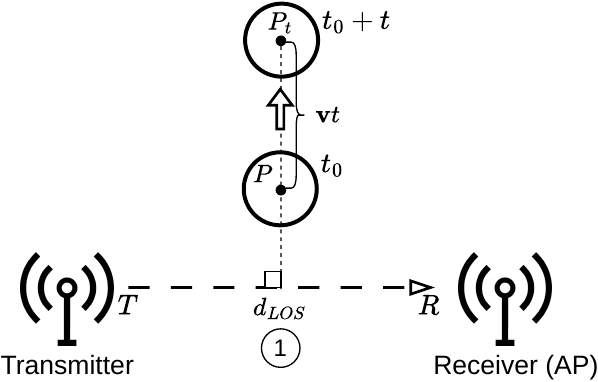}
    }
    \hfill
    \subfloat[$T\!\rightarrow\!P\!\rightarrow\!R$\label{fig:geo:b}]{%
        \includegraphics[width=0.48\linewidth]{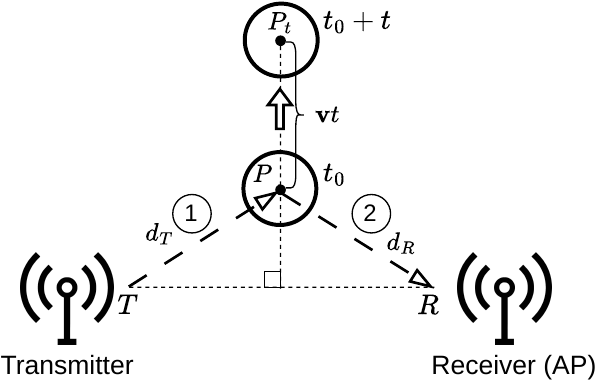}
    }\\[1ex]
    \subfloat[$T\!\rightarrow\!S\!\rightarrow\!P\!\rightarrow\!R$\label{fig:geo:c}]{%
        \includegraphics[width=0.48\linewidth]{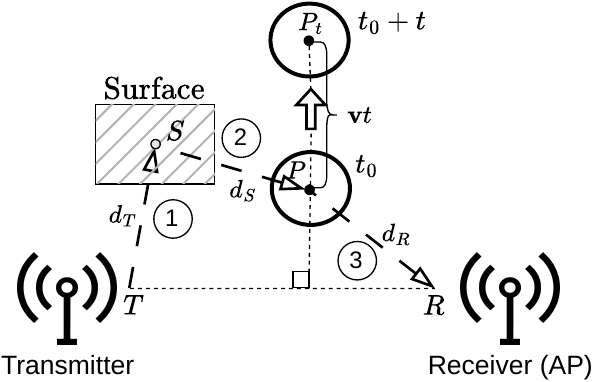}
    }
    \hfill
    \subfloat[$T\!\rightarrow P\!\rightarrow\!S\!\rightarrow\!R$\label{fig:geo:d}]{%
        \includegraphics[width=0.48\linewidth]{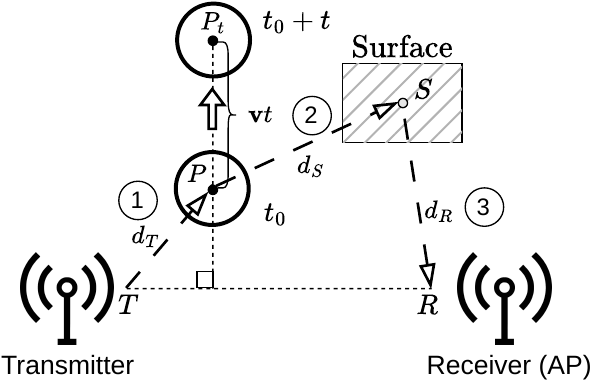}
    }
    \caption{The geometric configuration of signal propagation between a Wi-Fi transmitter \(T\) and receiver \(R\), considering a scenario in which a point \(P\) moves vertically with constant velocity \(\mathbf{v}\) over a time interval \( t\). Each subfigure illustrates a distinct propagation path:  
(a) the direct LOS path,  
(b) a path that reflects off the moving point before reaching the receiver,  
(c) a path that reflects off a surface \(S\) before interacting with the moving point, and  
(d) a path that reflects off a surface after the signal has interacted with the moving point.}
    \label{figure:geometry}
\end{figure}

\revised{The Wi-Fi signal can reach the receiver through several intuitive routes, as shown in Fig.~\ref{figure:geometry}. In this figure, \(T\) and \(R\) denote the transmitter and receiver, respectively, \(S\) denotes a static reflector such as a wall, desk, or chair, and \(P\) denotes the moving point, such as a hand or body part. The initial and displaced positions of the moving point are denoted by \(P_0=(x_P,y_P,z_P)\) and \(P_t=P_0+\mathbf{v}t\), respectively, where \(\mathbf{v}=(v_x,v_y,v_z)\) is the velocity vector of the moving point and \(t\) is the elapsed time. The simplest route is the direct line-of-sight path \(T\!\rightarrow\!R\), which does not interact with the moving point and mainly represents the static channel. A motion-sensitive route occurs when the signal reflects from the moving point \(P\), such as a hand or body part, before reaching the receiver, i.e., \(T\!\rightarrow\!P\!\rightarrow\!R\). In a real indoor environment, the signal may also interact with a static object \(S\), such as a wall, desk, or chair, either before or after it reaches the moving point. Examples include \(T\!\rightarrow\!S\!\rightarrow\!P\!\rightarrow\!R\), where the signal first reflects from the environment and then from the moving point, and \(T\!\rightarrow\!P\!\rightarrow\!S\!\rightarrow\!R\), where the signal first interacts with the moving point and then reflects from the environment. Higher-order paths can be interpreted similarly, but are typically weaker due to additional attenuation and can therefore even be neglected or omitted.}

\begin{figure}[!t]
	\centering
	\subfloat{%
		\includegraphics[width=0.35\linewidth]{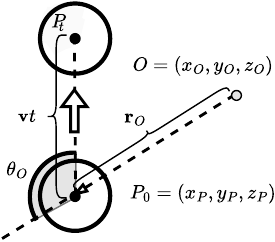}
	}
	\caption{Geometry of the distance change between a fixed observer \(O\) and a moving point \(P\), showing how the displacement \(\mathbf{v}  t\) affects the path length.
}
\label{figure:Proposition1}
\end{figure}

The propagation delay along each path is determined by how the arbitrary $3$-D displacement of a point alters the geometric length of its segments. Proposition~\ref{Proposition:3ddisplacement} formalizes this relationship for any fixed observer point, as illustrated in Fig.~\ref{figure:Proposition1}.

\begin{Proposition}
\label{Proposition:3ddisplacement}
\textit{For an arbitrary fixed observer $O=(x_O, y_O, z_O)$ and a moving point with initial position $P_0=(x_P, y_P, z_P)$ and velocity vector $\mathbf{v}$, define the relative position vector}
\[
\mathbf{r}_O = P_0 - O, \qquad d_O = \|\mathbf{r}_O\|, \qquad \theta_O = \angle(\mathbf{r}_O, \mathbf{v}).
\]
\textit{The change in distance $\Delta L_O$ between the observer and the moving point after displacement $\mathbf{v}t$ is approximated by}
\begin{equation}
\Delta L_O \approx \|\mathbf{v}\|\,t \cos\theta_O,
\qquad
\|\mathbf{v}\|\, t \ll d_O.
\label{eq:linear3d}
\end{equation}
\end{Proposition}
\revised{\begin{proof}
Please refer to Appendix~\hyperref[Appendix:A]{A}.
\end{proof}}

In a typical enclosed indoor environment, the distances of the surfaces are of the order of $d_O = 3$–$5$ meters, while typical human motion speeds are $\|\mathbf{v}\| \leq 1.5$ m/s over arbitrary short intervals $\, t \leq 0.2$ seconds. Hence, $\|\mathbf{v}\|\, t \leq 0.3$ m, which is significantly smaller than $d_O$, justifying the condition $\|\mathbf{v}\|\, t \ll d_O$.

The total change in path length for any multisegment trajectory is equal to the sum of the increments in the segments incident on $P$. Hence
\begin{align}
\Delta L_{T\rightarrow S\rightarrow P\rightarrow R}&=\Delta L_S+\Delta L_R,\\
\Delta L_{T\rightarrow P\rightarrow S\rightarrow R}&=\Delta L_T+\Delta L_S,\\
\Delta L_{T\rightarrow P\rightarrow R}&=\Delta L_T+\Delta L_R,
\end{align}
where each $\Delta L$ is given by  \eqref{eq:linear3d}. In general, for \(K\) distinct segments, the total change in path length due to the displacement of a moving point \(P\) can be approximated as
\begin{equation}
\Delta L_{\text{total}} \approx \sum_{i=1}^K \|\mathbf{v}\| \, t \, \cos\theta_{i},
\label{eq:delay_resolution}
\end{equation}
where \(\theta_{i}\) denotes the angle between the displacement vector \(\mathbf{v}\) and the direction of the \(i^{th}\) segment incident at \(P\). Introducing the equivalent observation angle,
\begin{equation}
\cos\theta_{\mathrm{eq}} = \frac{1}{K}\sum_{i=1}^K \cos\theta_{i},
\end{equation}
allows for a more compact expression
\begin{equation}
\Delta L_{\mathrm{total}} \approx K\,\|\mathbf{v}\|\,\, t\,\cos\theta_{\mathrm{eq}}.
\label{eq:total_delay_eq}
\end{equation}

This result implies that the cumulative effect of all incident path segments can be represented by a single virtual path component originating from an equivalent observer located such that \(\theta_O = \theta_{\mathrm{eq}}\), with a scaling factor of \(K\). Consequently, any composite propagation scenario involving the transmitter, receiver, environmental reflectors, and a moving point in $3$-D space can be modeled using the segment law of Proposition~1 to estimate the total variation of the length of the path.

The total propagation delay introduced by the equivalent delay change from all contributing paths is given by
\begin{equation}
\Delta \tau_{\mathrm{total}} = \frac{\Delta L_{\mathrm{total}}}{c},
\label{eq:delay_def}
\end{equation}
where \(c\) is the speed of light. Substituting from~\eqref{eq:total_delay_eq}, gives
\begin{equation}
\Delta \tau_{\mathrm{total}} \approx \frac{K\,\|\mathbf{v}\|\,\, t\,\cos\theta_{\mathrm{eq}}}{c}.
\label{eq:delay_total_compact}
\end{equation}

This delay variation directly affects the phase of each path component in the channel response. Assuming a carrier frequency \(f_c\), the resulting phase shift due to the delay change is
\begin{equation}
\Delta \phi = -2\pi f_c \Delta \tau_{\mathrm{total}}.
\end{equation}
Accordingly, the CSI at time \(s + \, t\) can be obtained from the response at time \(s\) as
\begin{equation}
H(s + \, t) = H(s) \, e^{-j 2\pi f_c \Delta \tau_{\mathrm{total}}}.
\label{eq:channel_evolution}
\end{equation}

This expression reveals that small displacements of the moving point induce time-varying phase shifts in the CSI, called Doppler shifts, governed by the projected velocity component along the effective observation direction. As such, tracking the evolution of the CSI phase over time enables inference of motion-induced path changes in enclosed environments.

\subsection{Multipath CSI Decomposition}

\revised{At sub-$7$\,GHz frequencies, the delay resolution of Wi-Fi is limited by the available bandwidth. With typical channel bandwidths of $20$--$80$\,MHz, the CSI can only separate multipath components up to a finite delay resolution, corresponding to several meters of propagation-length difference. Therefore, physical paths whose lengths differ by less than this resolution are grouped into the same resolvable delay component, or delay bin. Depending on the number of available subcarriers, the delay-domain CSI contains several such delay bins. Each delay bin consequently aggregates multiple physical paths that interact with the moving point through different reflection geometries.}

\revised{Although the physical paths within each delay bin cannot be isolated, the bins are generally not identical. Each delay bin contains the coherent superposition of all paths whose total propagation delays fall within the corresponding delay interval. Because these path sets differ from one delay bin to another, their reflection geometries and interaction angles with the moving object can also differ statistically. Transforming CSI into the delay domain therefore partitions the composite multipath response into several delay-indexed components, each representing a different unresolved mixture of propagation paths. These components do not provide clean directional observations. Instead, each delay bin can be interpreted as a virtual camera whose effective observation quality depends on how concentrated or dispersed the interaction angles are within that bin. Bins dominated by a narrow range of angles provide sharper motion-induced signatures, whereas bins containing highly diverse angles produce signatures that are mixed and blurred due to the superposition of various path contributions with different interaction geometries.

This structured diversity is more informative for activity recognition than the fully aggregated multipath response in the raw CSI. Since different delay bins contain different statistical mixtures of paths, they provide complementary observations of the same motion. Together, these observations make motion-induced Doppler variations more distinguishable while reducing the masking effect of strong static reflections that dominate the aggregate CSI. To formalize this decomposition, consider a scenario with \(L\) propagation paths where a point moves with three-dimensional velocity \(\mathbf{v}\). The CSI observed at the \(n^{\text{th}}\) subcarrier and time \(s + t\) can be expressed as}

\begin{equation}
\begin{aligned}
H_{n}(s\, +\, & t)
 = \sum_{i=0}^{L-1}
    \underbrace{%
      \exp\bigl(-j2\pi f_{n}\,\tau_{i}\bigr)
    }_{\text{Static path phase shift}} \\[2mm]
&\quad \times
    \underbrace{%
      \iint_{\mathbf{r}\in\mathbb{S}^2}
        \beta_{i}(\mathbf{r})\,
        \exp\bigl(j2\pi \,t \,f_{d}(s;\mathbf{r})\,\bigr)
        \,p_{\Omega}^{i}(\mathbf{r})\,\mathrm{d}\mathbf{r}
    }_{\text{Motion-induced Doppler contributions}},
\end{aligned}
\label{eq:CSI}
\end{equation}

\noindent
where \(f_n\) is the frequency of the \(n^{th}\) subcarrier, defined as
\begin{equation}
f_n 
= f_c + \left(n - \tfrac{N}{2}\right) \Delta f,
\quad 0 \le n \le N-1,
\label{eq:subcarrier_freq}
\end{equation}

\noindent
and \(\Delta f\) is the subcarrier spacing, calculated as the ratio of the system bandwidth to the number of subcarriers, \(N\). The parameter \(\tau_i\) denotes the static delay associated with the \(i^{\text{th}}\) path, such as the line-of-sight (LOS) component or other fixed reflections that do not interact with the moving point. The term \(f_d(s;\mathbf{r})\) represents the Doppler shift caused by motion along the unit direction \(\mathbf{r}\), and is given by

\begin{equation}
\begin{aligned}
f_d(s;\mathbf{r}) = \frac{f_c}{c} \|\mathbf{v}(s)\| \cos\theta_{\mathbf{v},\mathbf{r}}
\end{aligned}
\label{eq:doppler_shift}
\end{equation}

\noindent
where \(\theta_{\mathbf{v}, \mathbf{r}}\) denotes the angle between the unit direction vector \(\mathbf{r}\), with \(\|\mathbf{r}\|=1\), and the velocity vector \(\mathbf{v}\). Without loss of generality, \(\mathbf{r}\) is assumed to be a unit direction vector drawn from the angular probability distribution \(p_{\Omega}^{i}(\mathbf{r})\), representing the distribution of effective interaction directions around the moving point, such as the hand.

\revised{A directional distribution can characterize the path-direction statistics either within an individual delay bin, representing the unresolved path directions aggregated in that bin, or across delay bins, representing the set of dominant directions associated with different delay components. In a fully isotropic scattering environment, the direction of each path, \(\mathbf{r}\), is uniformly distributed on the unit sphere \(\mathbb{S}^2\). However, in real environments, scattering often deviates from perfect isotropy; clusters may predominantly appear in one quadrant or near the ground or ceiling. In such cases, \(\mathbf{r}\) follows a non-uniform distribution on \(\mathbb{S}^2\), which can be modeled by a \textit{von Mises--Fisher} (vMF) distribution~\cite{turbic2024correlation} on the sphere that concentrates directions around a specific mean direction \(\mathbf{m} \in \mathbb{S}^2\) as follows}

\begin{equation}
p_{\Omega}(\mathbf{r} \mid \mathbf{m}, \kappa)
\;=\;
\frac{\kappa}{4\pi\,\sinh(\kappa)}
\,\exp\!\bigl(\kappa\,\mathbf{r}\cdot \mathbf{m}\bigr),
\label{eq:distribution}
\end{equation}

\begin{figure}[!t]
	\centering
	\subfloat{%
		\includegraphics[width=0.8\linewidth]{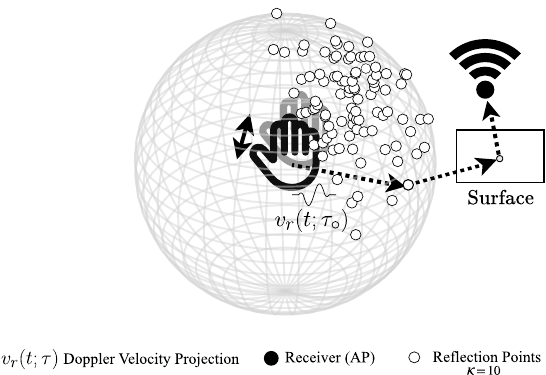}
	}
	\caption{\revised{Environmental reflection points are drawn from von Mises--Fisher distributions on \(\mathbb{S}^2\) with a concentration parameter \(\kappa=10\). Each point represents a reflection component that captures hand motion from a distinct angle and reaches the AP with an average delay of $\tau_\circ$. All directions are defined relative to the hand, which is located at the center of the sphere.}
}
\label{figure:sphere}
\end{figure}

\noindent
\revised{where \(\kappa > 0\) is the concentration parameter and \(\mathbf{m} \in \mathbb{S}^2\) is the mean direction of the scatterers. When \(\kappa=0\), the distribution reduces to the isotropic, uniform distribution on \(\mathbb{S}^2\); larger values of \(\kappa\) indicate stronger concentration around \(\mathbf{m}\). Fig.~\ref{figure:sphere} illustrates an example in which reflection points sampled from a vMF distribution with \(\kappa = 10\) observe hand motion from different directions. More generally, when multiple dominant scattering clusters exist, the corresponding angular distribution can be represented by a mixture of vMF distributions, providing a more flexible model for discrete and multimodal multipath directions. In this work, this distributional model is used only to provide intuition for how MORIC interprets multipath components as Doppler velocity projections from different observation directions. The vMF parameters are not fitted, estimated, or used during training.}

To decompose CSI into multipath components corresponding to distinct propagation delays, the $N$-point inverse discrete Fourier transform (IDFT) is applied with respect to the subcarrier index \(n\), as follows

\begin{equation}
\begin{aligned}
h(s+t;\tau) 
&= \mathcal{F}^{-1}\{H_n(s+t)\} \\[2mm]
&= \frac{1}{N} \sum_{n=0}^{N-1} H_n(s+t)\,
e^{j2\pi n \Delta f \tau}\,.
\end{aligned}
\label{eq:idft_direct}
\end{equation}

Substituting~\eqref{eq:CSI} into~\eqref{eq:idft_direct}, and absorbing the \(n\)-independent phase factor \(e^{-j2\pi (f_c-N\Delta f/2)\tau_i}\) into the complex path gain \(\beta_i(\mathbf{r})\), yields the delay-domain decomposition
\begin{equation}
\begin{aligned}
h(s+t;\tau)
&= \sum_{i=0}^{L-1}
\iint_{\mathbf{r} \in \mathbb{S}^2}
\beta_i(\mathbf{r})
\exp\Bigl(j2\pi t f_d(s;\mathbf{r})\Bigr) \\[2mm]
&\quad \times
\left[
\frac{1}{N} \sum_{n=0}^{N-1}
\exp\bigl(-j2\pi n \Delta f (\tau_i - \tau)\bigr)
\right]
p_{\Omega}^i(\mathbf{r}) \,\mathrm{d}\mathbf{r}\,.
\end{aligned}
\label{eq:idft_substitution_prob}
\end{equation}

The inner summation forms a finite geometric series that can be expressed as

\begin{equation}
\frac{1}{N} \sum_{n=0}^{N-1} \exp\left(-j 2\pi n \Delta f (\tau_i - \tau)\right)
= \frac{1}{N} \cdot \frac{1 - e^{-j2\pi N \Delta f (\tau_i - \tau)}}{1 - e^{-j2\pi \Delta f (\tau_i - \tau)}},
\end{equation}
\noindent
which simplifies into the Dirichlet kernel defined as

\begin{equation}
\begin{aligned}
\mathcal{D}_N(\tau - \tau_i)
&= \mathrm{sinc}_N\bigl(\Delta f(\tau - \tau_i)\bigr) \\
&\quad \times \exp\Bigl(j\pi(N-1)\Delta f (\tau - \tau_i)\Bigr),
\end{aligned}
\label{eq:sinc_dirichlet}
\end{equation}

\noindent
where the windowed \textit{sinc} function is given by

\[
\mathrm{sinc}_N(x) = \frac{\sin(\pi N x)}{N \sin(\pi x)}.
\]

Using this, the complete delay domain representation becomes

\begin{equation}
\begin{aligned}
h(s + t;\tau)
= &\sum_{i=0}^{L-1}
\mathrm{sinc}_N\bigl(\Delta f(\tau - \tau_i)\bigr) \\
&\times 
\exp\Bigl(j\pi(N-1)\Delta f (\tau - \tau_i)\Bigr) \\
&\times 
\iint_{\mathbf{r} \in \mathbb{S}^2}
\beta_i(\mathbf{r})\,
\exp\Bigl(j2\pi \, t f_d(s;\mathbf{r})\Bigr)\,
p_{\Omega}^{i}(\mathbf{r})\mathrm{d}\mathbf{r}.
\end{aligned}
\label{eq:idft_final_with_prob}
\end{equation}

The sinc envelope \(\mathrm{sinc}_N(\Delta f(\tau - \tau_i))\) characterizes the spectral leakage and temporal spread of the energy of each path in adjacent delay bins. The width of the mainlobe of this sinc function determines the system’s ability to resolve paths in the delay domain and is given by
\[
\Delta \tau_{min} = \frac{1}{N \Delta f} = \frac{1}{\mathrm{BW}},
\]
where \(\mathrm{BW} = N \Delta f\) denotes the total bandwidth of the system. This value defines the delay resolution, indicating the minimum distinguishable time difference between two multipath components. Specifically, two paths with delays \(\tau_i\) and \(\tau_j\) are considered resolvable if the absolute difference \(|\tau_i - \tau_j|\) exceeds \(\Delta \tau_{min}\). The corresponding minimum spatial resolution, or the ability to distinguish path lengths, follows as

\begin{equation}
\Delta d_{\min} = c \, \Delta \tau_{\min} = \frac{c}{\mathrm{BW}},
\label{eq:distance_resolution}
\end{equation}

\noindent
where \( c \) is the speed of light.

Although in practice there may exist \(L > N\) physical propagation paths, the IDFT yields only \(N\) resolvable delay bins centered at

\[
\tau_i = \frac{i}{N\Delta f},
\quad \text{for } i = 0, \dots, N-1.
\]

Therefore, based on the discrete delay value \(\tau_i\), CSI can be decomposed into \(N\) delay-specific multipath components. Regardless of the magnitude of \( \beta_i(\mathbf{r})\), each component provides a unique velocity measurement. Direct analysis of raw CSI often conceals motion contributions from weaker paths, as the dominant LOS component \(\lvert \beta_0 \rvert\) overshadows them. However, isolating the contribution of each delay bin allows for a more detailed characterization of motion by leveraging the spatial diversity inherent in the multipath structure.

Let \(\{\tau_0, \tau_1, \dots, \tau_{N-1}\}\) denote the set of discrete delays corresponding to the \(N\) resolvable components. The collection

\[
\mathcal{H}_{\mathcal{P}}(s)
=
\bigl\{
h(s;\tau_0),\,h(s;\tau_1),\,\dots,\,h(s;\tau_{N-1})
\bigr\}
\]
\noindent
is defined to capture the contribution of each multipath component to the overall channel as a function of delay \(\tau_i\). For notational simplicity, time is indexed by the discrete-time index $s$. Each bin \(\tau_i \pm \Delta \tau_{\min}/2\) can accumulate energy from one or more unresolved physical paths, with relative contributions determined by the associated angular distribution \(p_{\Omega}^i(\mathbf{r})\). As suggested by~(\ref{eq:total_delay_eq}), each delay bin can be interpreted as representing a superposition of multiple paths with similar propagation delays. The aggregate contribution of these paths introduces an unknown amplitude scaling, denoted by \(K\), which captures the combined effect of path gains and the number of contributing paths. Due to the subsequent normalization steps in the proposed method, each component \(h(s; \tau_i)\) is ultimately treated as a function of the velocity vector \(\mathbf{v}(s)\) and the equivalent observation angle \(\theta_{\mathrm{eq}}\), abstracting away the unknown scaling and path count.

\subsection{Doppler Velocity Information in Multipath Components}
\revised{Each delay-domain component \(h(s;\tau_i) \in \mathcal{H}_{\mathcal{P}}(s)\) contains motion-dependent information about the velocity vector \(\mathbf{v}(s)\) of the moving point. In general, because a delay bin aggregates multiple paths with different propagation directions, this information appears as a nonlinear mixture of velocity projections through the phase evolution of \(h(s;\tau_i)\). When the reflectors contributing to the delay bin are concentrated around a dominant direction, this mixture simplifies, and the phase dynamics directly reveal an approximately linear projection of \(\mathbf{v}(s)\) along that direction. The following propositions and formulations formalize this connection between delay-bin phase dynamics and the underlying motion velocity.}

\revised{To begin with, we need to reformulate the phase evolution in a way that avoids directly working with the phase operator \(\angle(\cdot)\). This is because the motion and velocity information carried by \(h(s;\tau_i)\) is embedded in its phase, but the phase operator is nonlinear and inconvenient to manipulate analytically. To address this, we use a standard identity from instantaneous frequency analysis~\cite{boashash1992estimating} that relates the phase derivative of a complex signal to its normalized complex derivative.}

\begin{Proposition}
\label{Proposition:phase_derivative}
\textit{Let \( h(s) = A(s) e^{j\zeta(s)} \in \mathbb{C} \), with \( A(s) > 0 \) and \( \zeta(s) \in \mathbb{R} \). Then,
\begin{equation}
\frac{d}{ds}\, \zeta(s)
\;=\;
\Im\left\{ \frac{\dot{h}(s)}{h(s)} \right\}.
\end{equation}
where \(\Im{\cdot}\) denotes the imaginary-part operator.}
\end{Proposition}

\revised{\begin{proof}
The proof is available in~\cite{boashash1992estimating}.
\end{proof}}

Now apply Proposition~\ref{Proposition:phase_derivative} to the multipath channel component at delay \(\tau_i\)

\begin{equation}
\begin{aligned}
\phi_i(s)
&= \angle h(s;\tau_i) \\
&= \angle\Biggl( \iint\limits_{\mathbb{S}^2} \beta_i(\mathbf{r}) \,
      e^{j 2\pi\,t\, f_c \frac{\mathbf{v}(s) \cdot \mathbf{r}}{c}} \, p_{\Omega}^i(\mathbf{r}) \, d\mathbf{r} \Biggr).
\end{aligned}
\label{eq:phi_integral}
\end{equation}

\noindent Assuming $v(s)$ approximately constant over the short local window, differentiating $\phi_i(s)$ with respect to $t$ and invoking the identity in Proposition~\ref{Proposition:phase_derivative} results in
\begin{equation}
\begin{aligned}
&\frac{d\phi_i(s)}{dt}
= \Im\left\{ \frac{\frac{d}{dt} h(s;\tau_i)}{h(s;\tau_i)} \right\} \\
&\quad = \frac{2\pi f_c}{c}
\ \Re \left\{ \frac{\iint_{\mathbb{S}^2} \beta_i(\mathbf{r}) \, (\mathbf{v}(s) \cdot \mathbf{r}) \,
      e^{j 2\pi\,t\, f_c \frac{\mathbf{v}(s) \cdot \mathbf{r}}{c}} \, p_{\Omega}^i(\mathbf{r}) \, d\mathbf{r}}{\iint_{\mathbb{S}^2} \beta_i(\mathbf{r}) \,
      e^{j 2\pi\,t\, f_c \frac{\mathbf{v}(s) \cdot \mathbf{r}}{c}} \, p_{\Omega}^i(\mathbf{r}) \, d\mathbf{r}} \right\}.
\end{aligned}
\label{eq:phase_derivative}
\end{equation}
\noindent
where \(\Im{\cdot}\) and \(\Re{\cdot}\) denote the imaginary-part and real-part operators, respectively.

\revised{Equation~\eqref{eq:phase_derivative} shows that the phase derivative of the delay-domain component is determined by the velocity-dependent term \(\mathbf{v}(s)\cdot\mathbf{r}\), indicating that it carries information about the performed activity. However, this relationship remains difficult to interpret because it is expressed as a ratio of angular integrals. The next propositions simplify this expression and provide a more interpretable form under certain conditions.}

\textit{Definition:} \textit{A function \( f \colon \mathbb{S}^2 \to \mathbb{C} \) is called smooth if it is infinitely differentiable on the sphere, i.e., \( f \in C^\infty(\mathbb{S}^2) \), where \( C^k(\mathbb{S}^2) \) denotes the space of functions whose derivatives up to order \( k \) exist and are continuous on \( \mathbb{S}^2 \).}

\begin{Proposition}
\label{Proposition:smooth_integrand}
\textit{Let $\mathbf{v}(s)\in\mathbb{R}^3$ be a fixed velocity vector.  
For constant parameters $f_c$, $c$, and $t$, define
\[
g(\mathbf{r})
=
\beta_i(\mathbf{r})\,
(\mathbf{v}(s)\!\cdot\!\mathbf{r})\,
\exp\!\Bigl(j \tfrac{2\pi f_c\,t}{c}\,\mathbf{v}(s)\!\cdot\!\mathbf{r}\Bigr)\,
\qquad \mathbf{r}\in\mathbb{S}^2,
\]
where $\beta_i:\mathbb{S}^2\!\to\!\mathbb{C}$ is smooth.}
\textit{Then $g$ is smooth on $\mathbb{S}^2$.}
\end{Proposition}

\revised{\begin{proof}
Please refer to Appendix~\hyperref[Appendix:B]{B}.
\end{proof}}

\begin{Proposition}
\label{Proposition:laplace_approx}
\textit{Let \( f(\mathbf{r}) \) be a smooth function on \( \mathbb{S}^2 \), and let \( p_{\kappa}(\mathbf{r}) \) be the vMF distribution with mean direction \(\mathbf{m}\) and concentration \(\kappa \gg 1\). Then,
\begin{equation}
\iint_{\mathbb{S}^2} f(\mathbf{r}) p_{\kappa}(\mathbf{r}) \, d\mathbf{r} = f(\mathbf{m}) + \mathcal{O}\left(\frac{1}{\kappa}\right).
\end{equation}}
\end{Proposition}

\revised{\begin{proof}
Please refer to Appendix~\hyperref[Appendix:C]{C}.
\end{proof}}

\begin{figure}[!t]
	\centering
	\subfloat{%
		\includegraphics[width=0.9\linewidth]{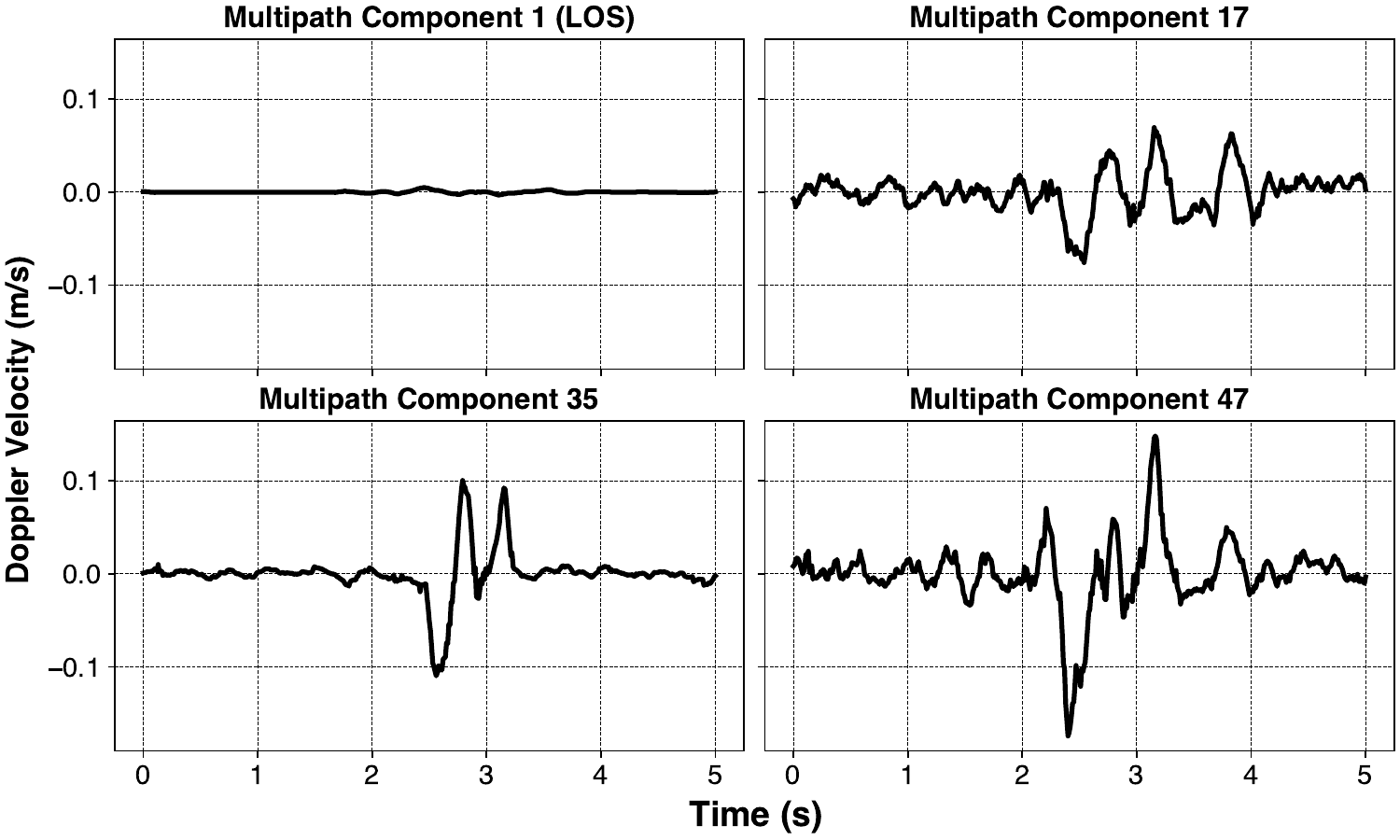}
	}
	\caption{\revised{Example Doppler velocity projections over time for four arbitrary delay bins, selected from \(52\) delay-domain components of a single antenna of a Wi-Fi AP, during a circle-drawing gesture. Because multipath propagation depends on the surrounding environment, the indices of non-LOS delay bins do not correspond to any fixed spatial order of observation directions. The first delay bin typically represents the LOS component and has the highest received power.}
}
\label{figure:doppler}
\end{figure}

By applying Proposition~\ref{Proposition:laplace_approx} to both the numerator and denominator of~\eqref{eq:phase_derivative}, whose integrands are smooth by Proposition~\ref{Proposition:smooth_integrand}, the expression is approximated as follows
\begin{equation}
\frac{d\phi_i(s)}{dt} \approx \frac{2\pi f_c}{c} \, \mathbf{v}(s) \cdot \mathbf{m}_i,
\label{eq:phase_laplace}
\end{equation}
where \(\mathbf{m}_i = \mathbb{E}_{p_{\Omega}^{i}}[\mathbf{r}]\) is the dominant or mean observation direction of the multipath component \(i\). As a result, the projected velocity on the multipath $i$ with delay $\tau_i$ can be described as

\begin{equation}
    v_r(s; \tau_i) = \frac{c}{2\pi f_c} \cdot \frac{d\phi_i(s)}{dt}.
\end{equation}

\noindent
where $v_r(s; \tau_i)$ is defined as 

\begin{equation}
    v_r(s; \tau_i) \delequal \mathbf{v}(s) \cdot \mathbf{m}_i = \mathbf{v}(s) \cdot \iint\limits_{\mathbb{S}^2} \mathbf{r} \, p_{\Omega}^{i}(\mathbf{r}) \, d\mathbf{r}.
\end{equation}

\noindent
which shows that, under a concentrated angular distribution within the \(i^{\text{th}}\) delay bin, \(v_r(s;\tau_i)\) provides an approximately linear projection of the actual three-dimensional velocity \(\mathbf{v}(s)\) along the corresponding observation direction. This linear projection interpretation is valid when \(\kappa \gg 1\), or equivalently, when the reflections contributing to the delay bin are concentrated around a dominant direction. Otherwise, \(v_r(s;\tau_i)\) remains a function of the actual velocity, but represents a mixed and blurred superposition of projections along multiple unknown directions. In this context, define the set
\[
\mathcal{V}_{\mathrm{r}}(s) =
\bigl\{
v_r(s;\tau_0),\,v_r(s;\tau_1),\,\dots,\,v_r(s;\tau_{N-1})
\bigr\},
\]
that collects projections of the true three-dimensional velocity vector \(\mathbf{v}(s)\) onto the directions associated with each multipath delay \(\tau_i\). \revised{This set consists of velocity projections that are implicitly functions of the concentration parameter $\kappa$ and the mean direction $\mathbf{m}_i$, which are typically unknown. However, it is not necessary to estimate these parameters, as the projections can be treated as distinct motion representations regardless of their specific values. Each multipath component \(h(s;\tau_i)\) captures motion from a distinct spatial angle, thus offering a rich and diverse description of the underlying movement. This diversity makes \(\mathcal{V}_{\mathrm{r}}(s)\) a powerful descriptor for HAR, as it encodes how motion is perceived through multiple propagation paths. In the proposed method, this set is used directly as input for training the classification model. Fig.~\ref{figure:doppler} illustrates an example in which four Doppler velocity projections extracted from a single CSI sample capture the same activity from different unknown observation directions. Indeed, these complementary projections are unordered and arise from unknown multipath-dependent directions. This motivates a classifier design that can aggregate them in an order-invariant manner and learn discriminative temporal signatures without relying on known Cartesian axes.}

\subsection{PSD-Based Doppler Velocity Estimation}

In multipath fading environments, the phase-derivative method for Doppler velocity estimation is often unreliable due to noise amplification and multipath interference. Differentiating the phase of \(h(s; \tau_i)\) amplifies noise and suffers from wrapping ambiguities caused by non-linear phase evolution, resulting in unstable velocity estimates. To address these issues and taking into account the angular distribution \(p_{\Omega}^{i}\), the power spectral density (PSD) has been shown to provide more accurate Doppler velocity estimates~\cite{gran2009adaptive}. PSD reveals Doppler shifts as spectral peaks in the frequency domain, effectively smoothing noise and avoiding phase discontinuities~\cite{narayanan2002doppler}.

In this context, the autocorrelation of \(h(s; \tau_i)\), denoted by \(R_h(\tau_i, t)\), is defined as
\begin{equation}
R_h(\tau_i, t) = \mathbb{E}\left[h^*(s; \tau_i)\, h(s\,+\,t; \tau_i)\right],
\end{equation}
\noindent
where \(\mathbb{E}(\cdot)\) denotes the expected value, and \(h^*\) is the complex conjugate of the channel \(h\). By substituting the expression for \(h(s; \tau_i)\), and assuming the scattering gains \(\beta_i(\mathbf{r})\) are uncorrelated across directions, the PSD \(S(f; \tau_i)\) is obtained via the Wiener--Khinchin theorem as the Fourier transform of the autocorrelation function as follows
\begin{equation}
S(f; \tau_i) = \int_{-\infty}^{\infty} R_h(\tau_i, t) \, e^{-j 2\pi f\,t} \, dt.
\end{equation}

This yields
\begin{equation}
\begin{aligned}
S(f; \tau_i) &= \int_{-\infty}^{\infty} \iint_{\mathbb{S}^2} |\beta_i(\mathbf{r})|^2 \,
\exp\left(j 2\pi\, t\frac{\mathbf{v}(s) \cdot \mathbf{r}}{\lambda}\right) \\
&\quad\quad\quad\quad \times p_{\Omega}^i(\mathbf{r}) \, e^{-j 2\pi f t} \, d\mathbf{r} \, dt \\
&= \iint_{\mathbb{S}^2} |\beta_i(\mathbf{r})|^2 \, p_{\Omega}^i(\mathbf{r}) \\
&\quad \times \left[ \int_{-\infty}^{\infty} \exp\left(j 2\pi\,t \left( \frac{\mathbf{v}(s) \cdot \mathbf{r}}{\lambda} - f \right) \right) dt \right] d\mathbf{r} \\
&= \mathbb{E}_{\mathbf{r} \sim p_{\Omega}^i} \left[ |\beta_i(\mathbf{r})|^2 \, \delta\left(f - \frac{\mathbf{v}(s) \cdot \mathbf{r}}{\lambda} \right) \right],
\end{aligned}
\end{equation}
\noindent where $\mathbb{E}$ denotes the expected value, defined as

\[
\mathbb{E}_{\mathbf{r} \sim p_{\Omega}^{i}}[\cdot] 
= \iint_{\mathbf{r} \in \mathbb{S}^2} (\cdot)\, p_{\Omega}^{i}(\mathbf{r})\, \mathrm{d}\mathbf{r}.
\]

The result indicates that the PSD is proportional to a probability density function on the Doppler frequency \(f\), determined by the projection of the spatial velocity vector \(\mathbf{v}(s)\) onto the scattering direction \(\mathbf{r}\), scaled by the wavelength \(\lambda\). Under the assumption of a sharply concentrated angular distribution, it can be shown that \(S(f; \tau_i)\) follows an approximately Gaussian profile~\cite{turbic2025doppler}. Assuming that both the angular distribution \(p_{\Omega}^{i}(\mathbf{r})\) and the gain profile \(\beta_i(\mathbf{r})\) are time-invariant, the projected Doppler velocity \(v_r(s; \tau_i)\) can be estimated as the frequency at which the PSD attains its maximum

\begin{equation}
v_r(s; \tau_i) = \lambda ~\underset{f}{\arg\max} \; S(f; \tau_i).
\end{equation}

This relationship arises because the PSD integral is maximized when the argument of the Dirac delta function is zero, that is, when
\[
f = \frac{\mathbf{v}(s) \cdot \mathbf{r}}{\lambda}.
\]

The PSD peak therefore reflects the projection of \(\mathbf{v}(s)\) with respect to \(\mathbf{r}\), modulated by the distribution \(p_{\Omega}^{i}(\mathbf{r})\). Under the assumption of a sharply concentrated vMF distribution with \(\kappa \gg 1\), Proposition~\ref{Proposition:laplace_approx} implies that the dominant contribution to the PSD arises from \(\mathbf{r} = \mathbf{m}_i\), yielding the peak frequency
\begin{equation}
    f^* = \frac{\mathbf{v}(s) \cdot \mathbf{m}_i}{\lambda},
\end{equation}

\noindent
where \(\beta_i(\mathbf{r})\) is assumed to be independent of time within the duration of the Welch window used for PSD estimation. This establishes a direct connection between the PSD peak and the projected velocity observed through the mean arrival direction \(\mathbf{m}_i\) of the multipath component \(h(s; \tau_i)\).

\revised{When the angular contribution is not sharply concentrated or contains multiple clusters, the PSD peak no longer corresponds to a single physical direction \(\mathbf{m}_i\). Instead, the estimated Doppler velocity should be interpreted as an effective projection that summarizes the dominant motion-sensitive scattering contributions within that delay bin.}

\subsection{\texorpdfstring{\revised{CSI Delay-Doppler Pre-processing}}{CSI Delay-Doppler Pre-processing}}
\revised{Before using \(\mathcal{V}_{\mathrm{r}}(s)\) for human activity classification, the extracted delay-Doppler representations must be preprocessed to reduce discontinuities and suppress unreliable components.} To mitigate jumps and spikes introduced during the transformation of CSI to Doppler velocity vectors, a Hampel filter \cite{pearson2016generalized} is applied to suppress these discontinuities while preserving the integrity of the original signal. The filter is applied independently to the real and imaginary parts of the channel components $h(s;\tau_i)$ for each delay \(\tau_i\). By identifying and replacing outliers based on a local median within a sliding window, the Hampel filter effectively removes abnormal fluctuations without significantly distorting the underlying signal. This approach ensures that activity-related information, which is critical for accurate analysis, is retained.

To account for the fact that some extracted multipath components and Doppler velocity projections may be dominated by noise or static background effects, which can obscure motion-induced patterns, these unreliable components should be suppressed. Otherwise, they may degrade model performance by introducing irrelevant or misleading information. To address this, the signal-to-noise ratio (SNR) is calculated for each extracted Doppler velocity projection. Assuming motion occurs near the center of the signal window, the SNR for each delay bin \(\tau_i\) is defined as the ratio of the variance during the motion segment to the variance during the static segments, which are taken as the initial and final \(10\%\) of the window:

\begin{equation}
\mathrm{SNR}_{\mathrm{dB}}(\tau_i) = 
10 \log_{10} \left[
\frac{
    \mathrm{Var}\left(v_r(s; \tau_i)\big|_{s \in \mathcal{T}_{\mathrm{motion}}}\right)
}{
    \mathrm{Var}\left(v_r(s; \tau_i)\big|_{s \in \mathcal{T}_{\mathrm{static}}}\right)
}
\right]
\label{eq:snr}
\end{equation}

\noindent where \(\mathcal{T}_{\mathrm{motion}}\) denotes the middle portion of the signal window and \(\mathcal{T}_{\mathrm{static}}\) represents the beginning and end segments. Components with \(\mathrm{SNR}_{\mathrm{dB}}(\tau_i) \leq 2\,\mathrm{dB}\) are discarded by replacing their corresponding velocity vectors with zeros. This filtering process encourages the model to focus on reliable motion-related patterns.

Moreover, the scale and amplitude of the Doppler velocity vectors depend on the path attenuation and the number of contributing path segments, represented by the scaling factor \(K\) in ~\eqref{eq:total_delay_eq}, which are generally unknown. This randomness introduces variability that is not informative for activity recognition. To mitigate its effect, each Doppler velocity vector, $v_r(s;\tau_i)$, is normalized to have zero mean and unit variance. This normalization ensures that the model focuses on the temporal pattern of motion rather than on the magnitude of the velocity.
\begin{figure*}[!t]
	\centering
	\subfloat{%
		\includegraphics[width=0.99\linewidth]{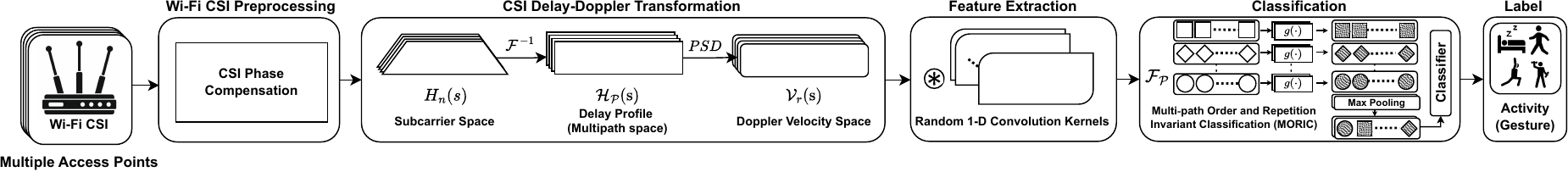}
	}
	\caption{The diagram illustrates the complete pipeline of the proposed Wi-Fi-based HAR method. First, Wi-Fi CSI signals are transformed into Doppler-velocity projections that capture movement velocity from multiple perspectives within the environment. Next, a feature-extraction stage, followed by a robust classifier designed to handle multipath randomness, recognizes the activity performed by the user.
}
\label{figure:method}
\end{figure*}

\subsection{Multipath Order and Repetition 
Invariant Classification (MORIC)}

The multipath velocity set \(\mathcal{V}_{\mathrm{r}}\) offers a comprehensive representation of the activity performed by capturing observations from multiple perspectives within the $3$-D environment. Each velocity projection \( v_r(s;\tau_i) \) constitutes a time-series whose characteristics vary according to the specific activity, making it suitable as input for subsequent feature extraction and classification methods. However, due to the stochastic nature of the environment, the exact viewpoint corresponding to each observation is unknown, resulting in variability and uncertainty in the ordering of elements within \(\mathcal{V}_{\mathrm{r}}\). \revised{Therefore, \(\mathcal{V}_{\mathrm{r}}\) should be treated as a set of motion representations rather than as a sequence with a reliable physical order. For example, if the dominant scattering contributions in one Doppler representation are concentrated around wall~A in one recording, a similar observation may appear at a different delay index in another recording when the effective propagation delays change. Meanwhile, nearby walls, furniture surfaces, or higher-order reflections may produce additional representations with related motion content. A classifier that depends on the input order could therefore learn a spurious association between a specific delay position and an activity, even though the same effective viewpoint may be shifted, permuted, or repeated across trials.} Moreover, unpredictable repetition of certain viewpoints can negatively impact classification performance, biasing predictions toward incorrect activities. \revised{The classifier must therefore be invariant to permutations of the elements of \(\mathcal{V}_{\mathrm{r}}\) and robust to repeated observations so that recognition depends on the temporal motion patterns contained in the representations, not on their arbitrary ordering.}

To address these challenges, this paper proposes \textit{MORIC}, a robust classification method that is invariant to the random ordering and repetition of velocity observations in \(\mathcal{V}_{\mathrm{r}}\). Fig.~\ref{figure:method} illustrates the different steps of the proposed method. \revised{The classification pipeline consists of two main components. The first component extracts temporal features independently from each Doppler velocity representation, so the sequential structure of each representation is modeled before any aggregation across multipath observations. The second component classifies the set of extracted representations using shared MLP heads, max-pooling over representations, and a final MLP classifier.} 

\subsubsection{Feature extraction}

\revised{The use of random convolution kernels for feature extraction has shown promising performance in many time-series classification tasks~\cite{lundy2021random, salehinejad2022litehar, salehinejad2023joint}, where unsupervised, training-free random convolutional features have been reported to perform competitively with, and in many cases outperform, state-of-the-art temporal models such as LSTMs, 1-D CNNs, and transformer-based classifiers for sequential and time-series classification tasks, including HAR, particularly when the available dataset is not very large. Random kernels help this task by acting as a large bank of temporal filters with diverse lengths, dilations, and biases, allowing local peaks, oscillations, transitions, and gesture-specific temporal patterns to be encoded from each Doppler velocity signal without training a high-capacity temporal model from scratch.}

Let $\mathbf{v}_r^{\tau_i}$ denote the velocity observation vector composed of $T$ samples of $v_r(s;\tau_i)$ for $s \in \{1,\dots,T\}$. The features are extracted from $\mathbf{v}_r^{\tau_i}$ using the Random Convolutional Kernel Transform method~\cite{dempster2020ROCKET}, which generates a large number of random convolutional kernels and applies them to the input signal to produce feature maps. Each kernel is constructed with a randomly selected length from $\{7, 9, 11\}$, weights sampled from a normal distribution, and dilation factors chosen as powers of two, enabling the capture of temporal patterns at multiple scales.

From each feature map, features are computed, including the maximum value, which identifies the strongest kernel response, and multiple proportions of positive values (PPV), which measure the frequency of positive activations by applying various bias terms to the convolution output. The resulting feature vector is denoted as $\mathbf{f}^{\tau_i} = (f_{1}^{\tau_i}, \dots, f_{D}^{\tau_i})$, where $D$ is the number of output features, typically on the order of thousands. For $N$ distinct delays, the set of feature vectors is given by $\mathcal{F}_P = \{\mathbf{f}^{\tau_0}, \mathbf{f}^{\tau_1}, \dots, \mathbf{f}^{\tau_{N-1}}\}$ and is used to train the model. All kernels and bias terms remain fixed across all velocity observation inputs, ensuring that the feature extraction process is invariant to data distribution shifts and yields a consistent feature set $\mathcal{F}_P$.

\subsubsection{Classification}
The extracted features from all velocity representation vectors are used to train a two-part neural network classifier. The first part comprises $K_C$ shared multi-layer perceptron (MLP) heads, each with two hidden layers, independently applied to every feature vector $\mathbf{f}^{\tau_i}$ to map the $D$-dimensional input to a lower-dimensional space of size $D'$. These MLPs are denoted by functions $g_k(\cdot)$ for $k = 1, 2, \dots, K_C$, and their outputs for each delay $\tau_i$ are given by
\[
\mathbf{f}^{\tau_i}_{\text{red},k} = g_k(\mathbf{f}^{\tau_i}), 
\quad \text{for } i = 0, 1, \dots, N-1,
\]
where each MLP performs non-linear dimensionality reduction, compressing $\mathbf{f}^{\tau_i}$ into a compact representation while capturing shared but diverse patterns across all velocity observation vectors.

For each head $k$, the resulting set of reduced feature vectors is denoted as
\[
\mathcal{F}_{\text{red},k}
= \{\mathbf{f}^{\tau_0}_{\text{red},k}, \mathbf{f}^{\tau_1}_{\text{red},k}, \dots, \mathbf{f}^{\tau_{N-1}}_{\text{red},k}\}.
\]
These are subsequently passed through a max-pooling layer that selects the maximum value across delays for each of the $D'$ dimensions. Formally, this operation is defined as
\[
\mathbf{f}^{\text{max}}_{k,d} 
= \max_{i \in \{0, 1, \dots, N-1\}} 
  \mathbf{f}^{\tau_i}_{\text{red},k,d},
\quad \text{for } d = 1, 2, \dots, D',
\]
where $\mathbf{f}^{\tau_i}_{\text{red},k,d}$ denotes the $d$-th component of the reduced feature vector from head $k$, and $\mathbf{f}^{\text{max}}_{k,d}$ is the corresponding max-pooled value. This step yields $K_C$ feature vectors of size $D'$, providing invariance to both the ordering and repetition of elements in $\mathcal{F}_P$. \revised{Max-pooling across delay-bin representations retains the strongest evidence for each learned feature, regardless of which multipath component produced it. Therefore, if a useful motion pattern appears in any reflection path, the classifier can use it without requiring that pattern to occur at a predetermined input position. Since the pooling operation keeps only the maximum activation in each feature dimension, repeated weak or redundant representations do not accumulate and are less likely to bias the final decision.}

Finally, the max-pooled feature vectors from all heads are concatenated and passed through a second MLP, denoted by $h(\cdot)$, to produce the class scores. A softmax function is applied to these scores to obtain the predicted class probabilities
\begin{equation}
\mathbf{p} 
= \operatorname{softmax}\left(h\left([\mathbf{f}^{\text{max}}_1 \| \mathbf{f}^{\text{max}}_2 \| \dots \| \mathbf{f}^{\text{max}}_{K_C}]\right)\right),
\end{equation}
where $\mathbf{p}$ represents the predicted probability distribution over the set of possible activity classes.

\subsection{\texorpdfstring{Calibration \revised{(optional)}}{Calibration (optional)}}

Hand motion patterns and their corresponding activity classifications may vary significantly across users, even for the same activity. This inter-user variability can lead to misaligned confidence estimates in the classifier predictions. To improve accuracy and ensure that the predicted probabilities \(\mathbf{p}\) better reflect user-specific likelihoods, optional Platt scaling~\cite{niculescu2005predicting} is applied as a post-processing calibration method.

The softmax outputs of the classifier often exhibit overconfidence, particularly when adapting to new users with limited data. Logit calibration addresses this by fitting a logistic regression model to the logits \(\mathbf{z} = h(\mathbf{f}^{\text{max}})\) using a small set of held-out calibration samples. The calibrated probabilities \(\mathbf{q}\) are computed for each class \(c \in \{1,\dots,C\}\), where \(C\) is the total number of activity classes, as
\[
q_c =
\frac{
\exp\left(\mathbf{w}_c^\top \mathbf{z} + b_c\right)
}{
\sum_{j=1}^{C}
\exp\left(\mathbf{w}_j^\top \mathbf{z} + b_j\right)
},
\]
where \(\mathbf{w}_c\) and \(b_c\) are the class-specific logistic-regression weight vector and bias for class \(c\), respectively. Equivalently, this can be written as
\[
\mathbf{q} = \mathrm{softmax}\left(\mathbf{W}\mathbf{z}+\mathbf{b}\right),
\]
where \(\mathbf{W}\in\mathbb{R}^{C\times C}\) and \(\mathbf{b}\in\mathbb{R}^{C}\) are optimized to minimize the negative log likelihood on the calibration data. Unlike temperature scaling, which uses a single global temperature parameter, logit calibration's class-specific parameters allow for finer adjustments to the decision boundaries, potentially improving both calibration and prediction accuracy for underrepresented classes. This calibration process preserves the hierarchical feature extraction of the model, via \(g_k(\cdot)\) and \(h(\cdot)\), while aligning its confidence estimates with empirical frequencies.

\section{Experiment}\label{section:experiment}

\subsection{Data}
\begin{figure}[!t]
	\centering
	\subfloat{%
		\includegraphics[width=0.8\linewidth]{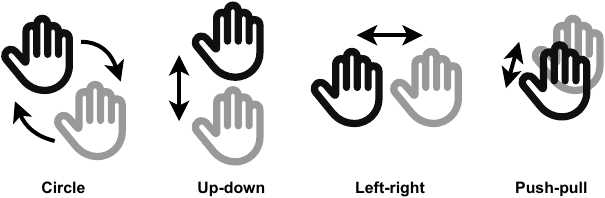}
	}
	\caption{Four hand activities performed by users in the UTHAMO dataset.
}
\label{figure:gestures}
\end{figure}
\begin{figure}[!t]
    \centering
    \newcommand{\uthamoactionimage}[1]{%
        \makebox[0.48\linewidth][c]{%
            \resizebox{0.48\linewidth}{0.36\linewidth}{\includegraphics{#1}}%
        }%
    }
    \subfloat[\normalfont\revised{Circle}\label{fig:uthamo_action_circle}]{%
        \uthamoactionimage{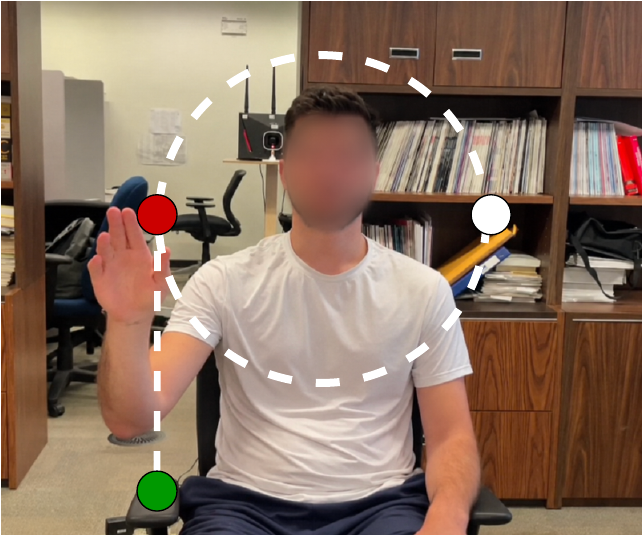}
    }
    \hfill
    \subfloat[\normalfont\revised{Up--down}\label{fig:uthamo_action_updown}]{%
        \uthamoactionimage{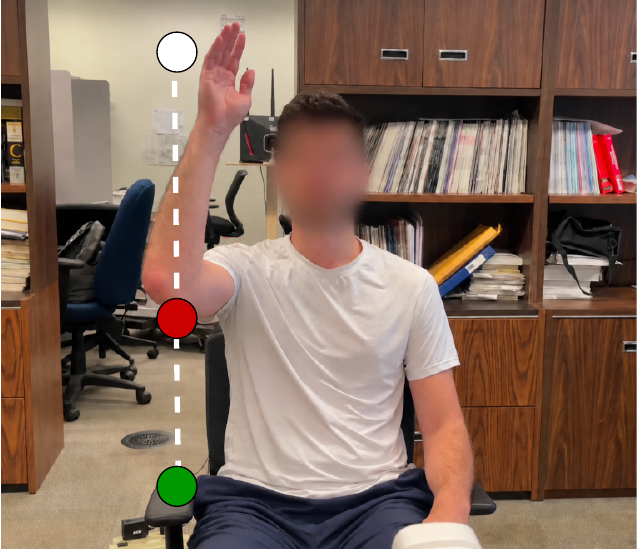}
    }\\[1ex]
    \subfloat[\normalfont\revised{Left--right}\label{fig:uthamo_action_leftright}]{%
        \uthamoactionimage{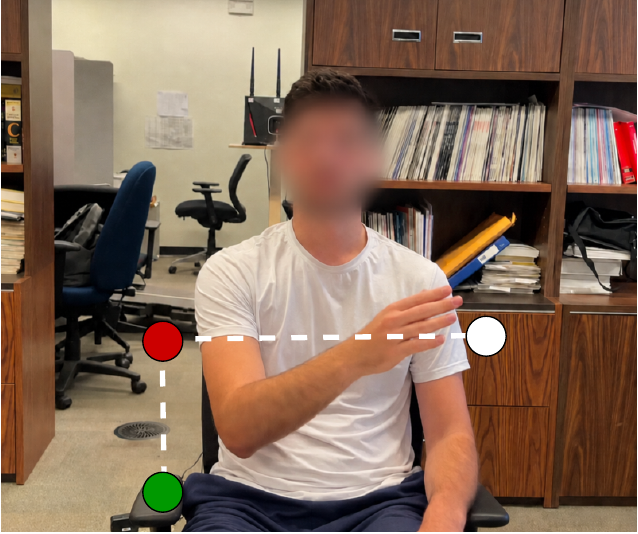}
    }
    \hfill
    \subfloat[\normalfont\revised{Push--pull}\label{fig:uthamo_action_pushpull}]{%
        \uthamoactionimage{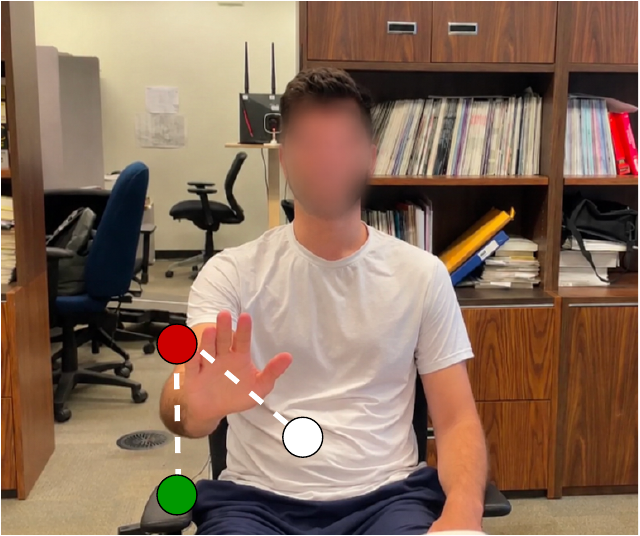}
    }
    \caption{\revised{Real-photo examples of the four human action classes in the UTHAMO dataset. The participant's hand initially rests on the chair handle, marked by the green point. To perform the activity, the participant moves the hand to the initial point, marked by the red point, and then starts the hand movement. The white dot indicates the middle of the performed action, and the dashed lines indicate the hand-movement trajectory.}}
    \label{figure:uthamo_action_photos}
\end{figure}
\begin{figure}[!t]
    \centering
    \subfloat[\normalfont\revised{Front view}\label{fig:uthamo_env_front}]{%
        \includegraphics[width=0.7\linewidth]{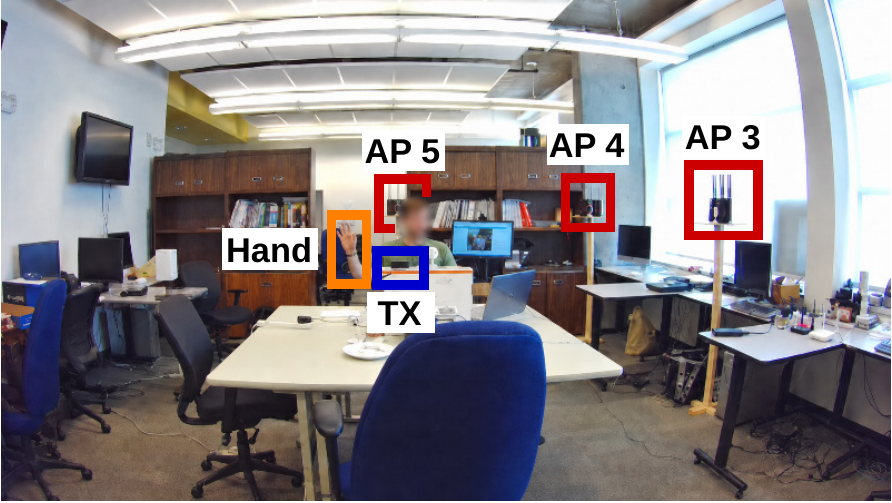}
    }\\[1ex]
    \subfloat[\normalfont\revised{Behind view}\label{fig:uthamo_env_behind}]{%
        \includegraphics[width=0.7\linewidth]{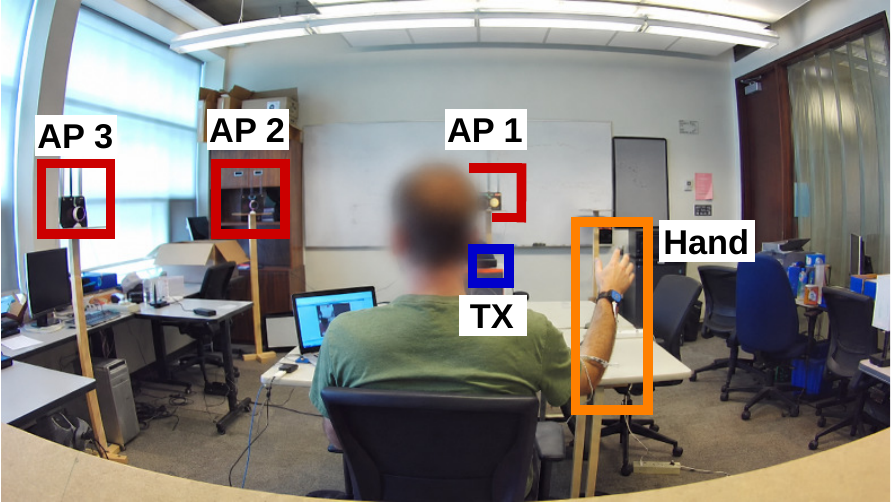}
    }
    \caption{\revised{Front and behind views of the UTHAMO data-collection environment.}}
    \label{figure:uthamo_environment_views}
\end{figure}

To evaluate the performance of the proposed method, in this work a dataset on hand motion, named (UTHAMO), has been collected\footnote{\revised{The UTHAMO dataset is publicly available at \url{https://ieee-dataport.org/documents/uthamo-multi-modal-wi-fi-csi-based-hand-motion-dataset-0}.}}. The hand motion dataset consists of CSI recordings corresponding to four right-hand gestures—\textit{circle}, \textit{left–right}, \textit{up–down}, and \textit{push–pull}—performed by six adult participants in a static indoor office environment of size \(6\,\text{m} \times 5.6\,\text{m}\). The gestures are visually shown in Fig.~\ref{figure:gestures}. \revised{Representative real-photo examples with annotated hand trajectories and views of the data-collection environment are shown in Fig.~\ref{figure:uthamo_action_photos} and Fig.~\ref{figure:uthamo_environment_views}, respectively.} All participants gave informed consent.
 The gestures are selected to have analogous two-dimensional projections of motion in 3-D space, allowing us to examine how effectively Wi-Fi-based HAR methods distinguish small-scale gestures that may appear similar from certain observation directions. For example, a \textit{push--pull} motion in the sagittal plane can produce projections that resemble \textit{up--down} or \textit{left--right} motions in the coronal plane, making the gestures difficult to separate when the effective observation direction is unknown.

\begin{figure}[!b]
        \centering		\includegraphics[width=0.74\linewidth]{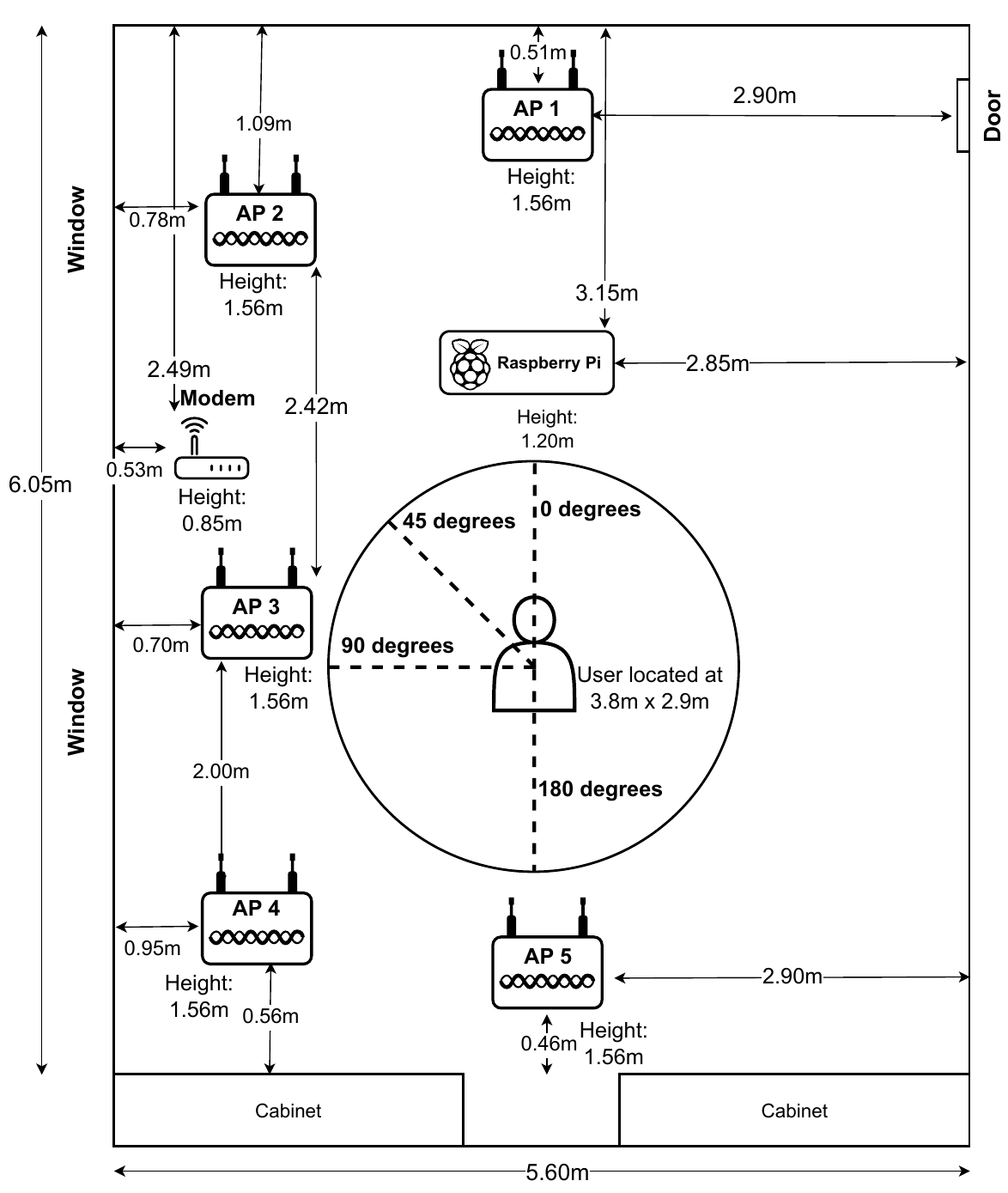}
	
	\caption{The detailed floor plan of the UTHAMO data collection setup.}
	\vspace{-0mm}
\label{figure:map}
\end{figure}

Data were collected at four body orientations (\(0^{\circ}\), \(45^{\circ}\), \(90^{\circ}\), and \(180^{\circ}\)) using five ASUS RT-AC86U access points (APs) positioned around the subject, as illustrated in Fig.~\ref{figure:map}. Each AP was equipped with three antennas and operated as a passive sniffer. CSI collection and packet sniffing were performed using the Nexmon CSI toolkit~\cite{nexmon}. \revised{The APs operated in monitor mode, passively intercepting packets transmitted by a Raspberry~Pi~3 Model~B+ over the \(2.4\,\text{GHz}\) Wi-Fi band at a rate of \(100\,\text{Hz}\).}

For each combination of body orientation and gesture, \(20\) trials were recorded, yielding \(1{,}920\) samples in total. Each trial consisted of a \(15\)-second rest period, a \(4\)-second video-guided gesture, a \(6\)-second post-gesture rest, and a \(20\)-second inter-trial pause. For each gesture, CSI collection covered a \(5\)-second window, starting \(0.5\) seconds before the gesture onset and ending \(0.5\) seconds after the \(4\)-second gesture.

Each antenna produced \(64\) subcarriers, of which \(52\) carried channel information and the remainder corresponded to pilot, guard, or null tones. Thus, each AP produced a complex-valued CSI matrix of size \(156 \times T\), where \(T\) is the number of temporal samples and equals \(500\) for a five-second recording.

\begin{table*}[!t]
\centering
\caption{Four-class hand-motion generalization accuracy (\%) on the UTHAMO dataset for orientation $180^\circ$. (Mean $\pm$ SD)}
\label{table:results}
\begin{adjustbox}{width=0.98\textwidth}
\begin{tabular}{ll|cccccc}
\toprule
\textbf{Method} & \textbf{Input Signal} 
  & \textbf{SAP $1$} & \textbf{SAP $2$} & \textbf{SAP $3$} & \textbf{SAP $4$} & \textbf{SAP $5$} & \textbf{All APs} \\
\midrule
AMAP~\cite{salehinejad2023joint}            & Raw CSI Magnitude      
  & 25.4\%\,$\pm$\,2.2\%  & 25.2\%\,$\pm$\,2.4\%  & 25.5\%\,$\pm$\,2.6\%  & 26.8\%\,$\pm$\,3.1\%  & 27.4\%\,$\pm$\,2.6\%  & 26.3\%\,$\pm$\,2.8\%  \\
\midrule
CMAP~\cite{salehinejad2023joint}            & Raw CSI Magnitude         
  & 25.2\%\,$\pm$\,2.3\%  & 27.2\%\,$\pm$\,2.3\%  & 26.0\%\,$\pm$\,2.7\%  & 25.9\%\,$\pm$\,2.5\%  & 27.5\%\,$\pm$\,2.6\%  & 28.1\%\,$\pm$\,2.9\%  \\
\midrule
CapsHAR~\cite{djogoHAR}                    & CSI Magnitude 
  & 28.2\%\,$\pm$\,4.2\%  & 29.1\%\,$\pm$\,3.4\%  & 27.9\%\,$\pm$\,3.7\%  & 27.8\%\,$\pm$\,4.5\%  & 33.4\%\,$\pm$\,4.2\%  & 33.7\%\,$\pm$\,6.1\%  \\

\midrule
CSI Ratio Model~\cite{wu2022wifi}          & CSI Ratio Phase       
  & 28.1\%\,$\pm$\,3.5\%  & 27.6\%\,$\pm$\,4.1\%  & 28.4\%\,$\pm$\,4.4\%  & 28.9\%\,$\pm$\,5.2\%  & 35.5\%\,$\pm$\,5.6\%  & 36.1\%\,$\pm$\,5.1\%  \\
\midrule
APNSS + APSC~\cite{102859317}              & Doppler Velocity 
  & 26.7\%\,$\pm$\,2.3\%  & 29.1\%\,$\pm$\,5.8\%  & 27.3\%\,$\pm$\,2.9\%  & 28.4\%\,$\pm$\,3.5\%  & 39.1\%\,$\pm$\,6.0\%  & -- \\
\midrule
SHARP~\cite{meneghello2022sharp}           & Doppler Spectrogram 
  & 28.7\%\,$\pm$\,3.8\%  & 27.3\%\,$\pm$\,3.1\%  & 29.2\%\,$\pm$\,4.5\%  & 28.9\%\,$\pm$\ 3.5\%  & 33.9\%\,$\pm$\,6.7\%  & 34.8\%\,$\pm$\,8.2\%  \\
\midrule
Ours (MORIC)                                & Multipath Doppler Velocities       
  & \textbf{39.0\%\,$\pm$\,7.3\%}  
  & \textbf{37.9\%\,$\pm$\,7.1\%} 
  & \textbf{39.4\%\,$\pm$\,7.8\%}
  & \textbf{47.7\%\,$\pm$\,9.8\%} 
  & \textbf{51.5\%\,$\pm$\,8.7\%} 
  & \textbf{56.3\%\,$\pm$\,9.1\%} \\
\bottomrule
\end{tabular}
\end{adjustbox}
\end{table*}

\subsection{Training and Test Procedure}
MORIC was implemented in PyTorch with the number of heads set to $K_C=2$, the number of kernels to $D=1{,}000$, a hidden layer size of $256$ for the first MLP, $D^\prime=128$ for the reduced feature dimension, and a cross-entropy loss with label smoothing of $0.1$. The model was trained with AdamW optimizer~\cite{loshchilov2017decoupled} with a learning rate of $1 \times 10^{-4}$, a batch size of $64$, and a maximum of $2500$ epochs. \revised{To assess the practical performance of the proposed method under cross-user generalization, a leave-one-subject-out (LOSO) cross-validation strategy was employed. In each iteration, the model was trained and validated using data from all participants except one, and then tested on the excluded participant. This process was repeated until each participant served once as the unseen test subject. To reduce overfitting, early stopping with a patience of \(200\) epochs was applied, and the model with the lowest validation loss was selected.}

\subsection{Results}

\subsubsection{Model Generalization to Unseen Users}

\begin{figure}[!b]
        \centering		\includegraphics[width=0.9\linewidth]{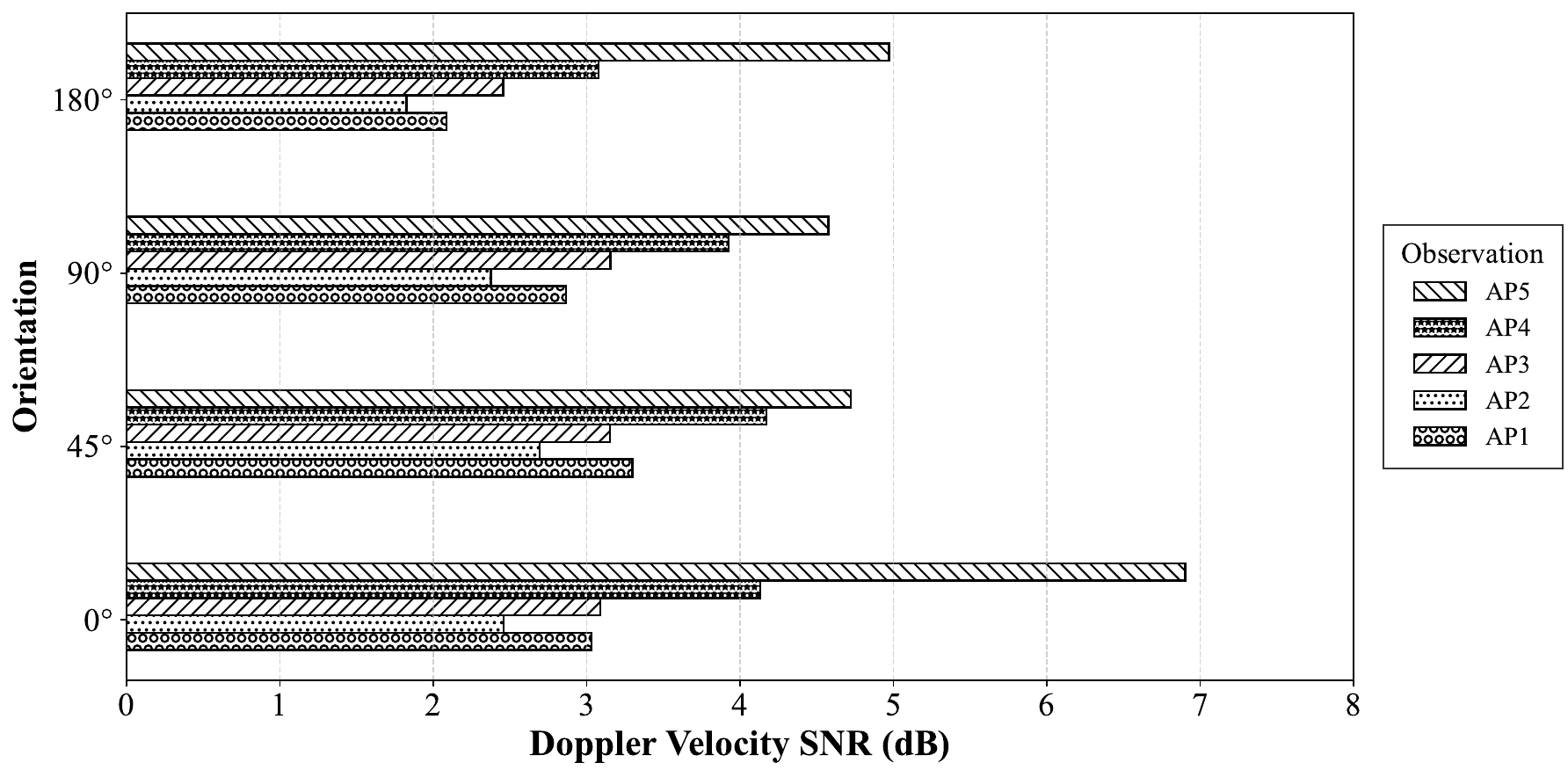}
	
\caption{Doppler velocity vector SNR in dB, computed using~\eqref{eq:snr} and aggregated by the median across all $N$ multipath components, for different APs and user orientations.}

	\vspace{-0mm}
\label{figure:snr}
\end{figure}

Table~\ref{table:results} presents the performance of the proposed method, MORIC, alongside other existing methods from the literature for Wi-Fi-based HAR on the UTHAMO dataset for orientation $180^\circ$. The evaluation considers two settings: one in which only a single AP (SAP) among the five available APs is used for HAR, and another in which the CSI from all APs is concatenated and used as input to MORIC and all other baselines. For four-class classification, the results show that the proposed method, MORIC, outperforms all other methods with an average accuracy of $56.3\%$ when all APs are used. Among the SAP scenarios, SAP~$5$ demonstrates significantly better generalization accuracy compared to the other SAPs, achieving a generalization accuracy of $51.5\%$. Table~\ref{table:avg_confusion} shows the confusion matrix for orientation $180^\circ$, indicating that the generalization accuracy for each class ranges from $50.0\%$ to $62.1\%$, corresponding to the \textit{push-pull} and \textit{up–down} gestures, respectively.

Traditional methods that directly utilize the magnitude or phase of raw CSI for model training perform poorly in generalization assessments. Approaches such as AMAP and CMAP~\cite{salehinejad2023joint} achieve accuracy levels close to the chance rate, indicating that while they may perform well under fixed settings with the same user, their performance degrades significantly in real-world scenarios. A key reason for this degradation is that, although the magnitude of the CSI exhibits activity-related patterns, it is highly sensitive to noise and environmental factors. CapsHAR~\cite{djogoHAR} slightly improves generalization utilizing a more advanced architecture and processing pipeline, achieving an accuracy of $33.7\%$. As shown in Table~\ref{table:results}, the CSI ratio model, which mitigates the effects of STO and SFO in the phase, followed by a random convolutional kernel transform, results in a modest improvement in the average accuracy, reaching $36.1\%$ when using the processed CSI phase. This highlights the capability of CSI phase information to offer additional features tied to activities. However, even with this advancement, the overall accuracy is still far from adequate for reliable use in real-world scenarios.

\begin{table}[!t]
\centering
\caption{Average confusion matrix (\%) across six subjects (orientation $180^\circ$, ALL APs)
}
\begin{tabular}{c|cccc}
\hline
 & \textbf{Circle} & \textbf{Left–right} & \textbf{Up–down} & \textbf{Push–pull} \\
\hline
\textbf{Circle}     & \textbf{56.4} & 10.5 & 20.8 & 12.3 \\
\textbf{Left–right} & 14.2 & \textbf{56.6} & 8.4 & 20.8 \\
\textbf{Up–down}    &  17.5 &  5.4 & \textbf{62.1} & 15.0 \\
\textbf{Push–pull}  & 9.1 & 23.4 & 17.5 & \textbf{50.0} \\
\hline
\end{tabular}
\label{table:avg_confusion}
\end{table}

Methods that utilize Doppler velocity have demonstrated more promising generalization performance. The APNSS$\,+\,$APSC approach~\cite{102859317}, which selects the best pair of antennas from a given AP and then extracts the corresponding Doppler velocity, achieves a higher average accuracy of $39.1\%$ compared to methods that directly use raw CSI or CSI ratio signals, indicating that Doppler velocity provides richer information about the movements performed. However, APSC is not designed to operate with multiple APs, which limits its effectiveness in environments where multiple APs are available.
The SHARP method~\cite{meneghello2022sharp} constructs a Doppler spectrogram from CSI data but still achieves a relatively low average accuracy of $34.8\%$ using all APs, based on the default parameters and settings from the original implementation. Although SHARP relies on the extraction of Doppler shifts present in the CSI, it derives a single resultant Doppler shift across all available multipaths in the environment. Consequently, unlike MORIC, SHARP is limited in its ability to distinguish fine-grained activities or gestures with similar motion characteristics.

\begin{figure*}[!t]
	\centering

		\includegraphics[width=1.0\linewidth]{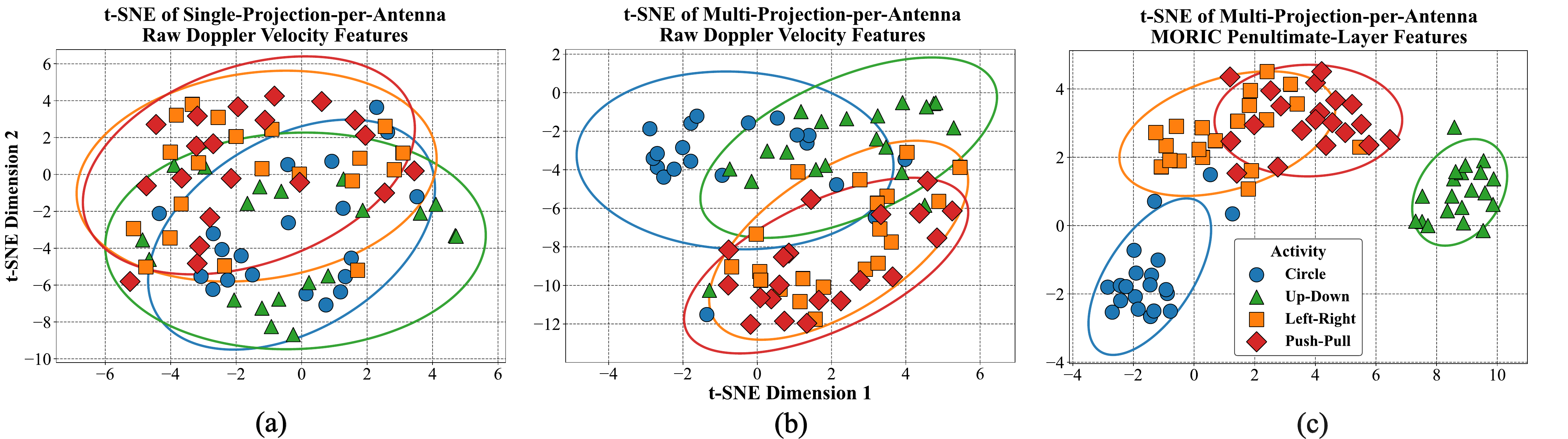}
	
		\caption{\revised{t-SNE visualizations of hand-motion representations for four activities from an example participant, using a single AP (AP~\(5\)) with three antennas as input. \textbf{(a)} Features extracted from one Doppler velocity projection per antenna exhibit poor class separability, highlighting the difficulty of fine-grained HAR with limited Doppler observations. \textbf{(b)} Features extracted from multiple Doppler velocity projections per antenna provide complementary motion perspectives and yield partial activity separability even before MORIC representation learning: Circle is reasonably separated from Push--Pull and Left--Right, whereas Circle and Up--Down still exhibit pronounced overlap, and Left--Right and Push--Pull remain nearly inseparable. \textbf{(c)} Penultimate-layer features from the proposed MORIC provide substantially improved activity-specific separation, although some overlap remains between Left--Right and Push--Pull. Colored boundaries denote covariance ellipses spanning two standard deviations for each activity. The t-SNE axes do not have direct physical or quantitative meaning and are used only for visualization.}}
		\label{figure:tsne_representations}
\end{figure*}

\revised{The t-SNE representation visualization in Fig.~\ref{figure:tsne_representations} provides a qualitative example of why multiple delay-separated Doppler projections are useful. Using only one Doppler velocity projection per antenna from AP~$5$ yields strongly mixed feature clusters, showing that a limited set of Doppler observations can leave fine-grained hand motions ambiguous. When multiple Doppler velocity projections are extracted per antenna, the same activity is observed from several multipath-dependent perspectives; for example, a circular motion may appear similar to a left--right or push--pull motion in one projection, while another projection can reveal complementary oscillatory or transverse motion cues. These complementary projections yield partial separability even before MORIC representation learning. The penultimate-layer MORIC features further show that the proposed model can unfold the ambiguity present in individual Doppler velocity projections and make the activities more distinguishable by aggregating unordered and potentially repeated projection-level evidence, reducing sensitivity to outlier projections, and extracting higher-level motion-discriminative features.
}

\subsubsection{Impact of Orientation and Geometry on Performance}
As illustrated in Fig.~\ref{figure:map}, the user is positioned between AP~$5$ and the Raspberry Pi device, which functions as the transmitter. This placement obstructs the LOS path to AP~$5$ and partially to AP~$4$. As a result, the higher average generalization accuracy for SAP~$5$ at orientation $180^\circ$, compared to other SAPs, can be attributed to the substantial attenuation of the LOS component, which typically carries limited motion-related information. This attenuation leads to a redistribution of signal power across non-line-of-sight (NLOS) paths, which are more sensitive to user motion and therefore enhance recognition performance. In this scenario, these alternative NLOS paths tend to exhibit a higher SNR and contribute more effectively to the received signal, enriching motion-related information captured through multipath components from various spatial perspectives. In this context, Fig.~\ref{figure:snr} confirms that, across all orientations, AP~$5$ consistently achieves significantly higher median SNR values for extracted Doppler velocity projections $v_r(s; \tau_i)$. The high performance of SAP~$5$ for orientation $180^\circ$ suggests that a single AP may be sufficient for HAR in real-world applications, provided that the AP and transmitter are positioned so that the user blocks the LOS path.

\begin{figure}[!b]
        \centering		\includegraphics[width=0.88\linewidth]{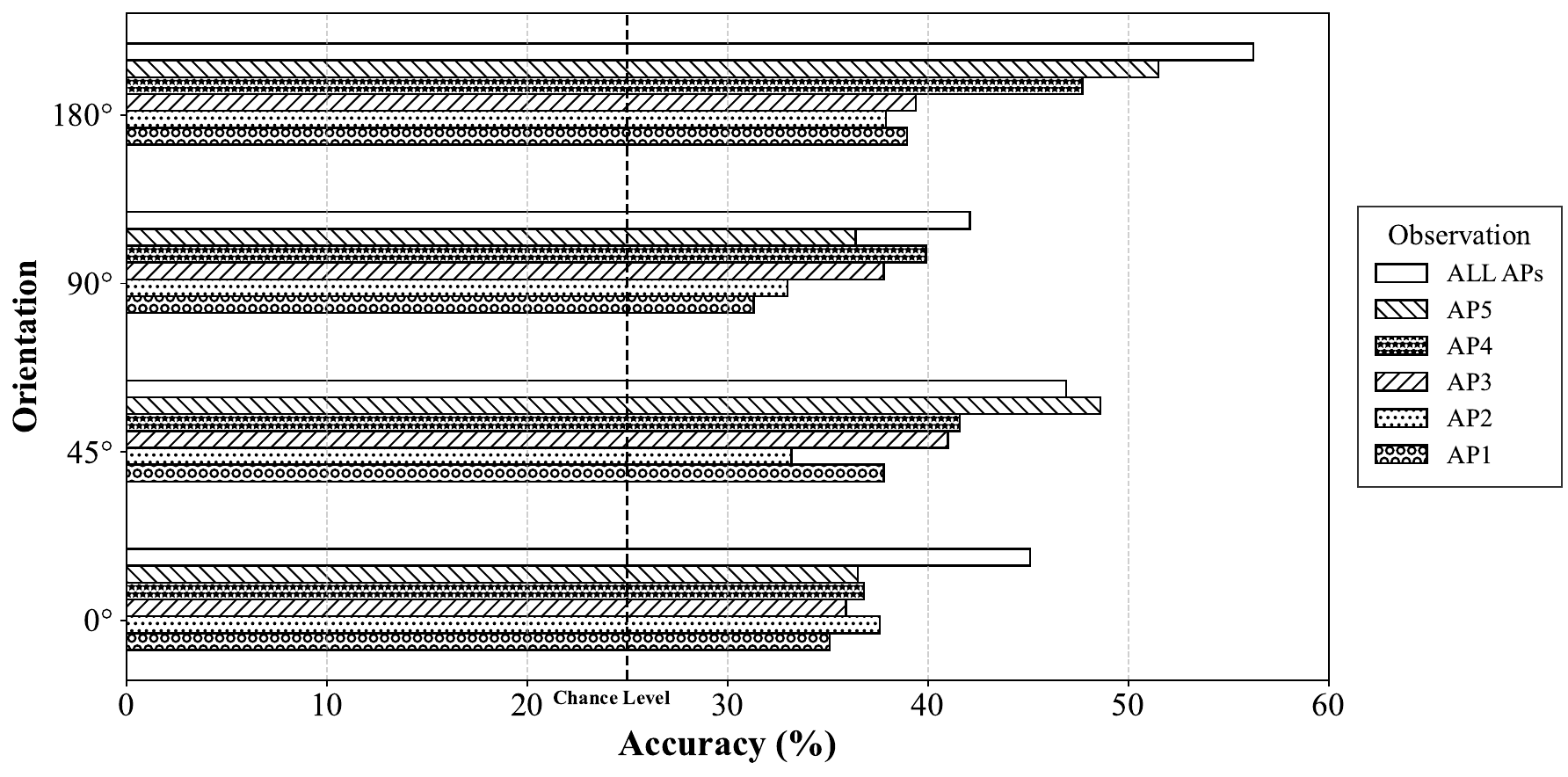}
	
	\caption{Generalization accuracy of MORIC to unseen users averaged over six participants for \textbf{four hand motion classes} across different user orientations and both single and multiple AP configurations.}
	\vspace{-0mm}
\label{figure:4classesHist}
\end{figure}

Fig.~\ref{figure:4classesHist} illustrates how generalization accuracy varies across different orientations and AP observations. Among the single APs, AP~$5$ and AP~$4$ consistently achieve the highest generalization accuracy in all orientations. Notably, SAP~$4$ performs best in orientation $90^\circ$, where the user more strongly intersects the LOS path. This highlights the critical influence of the geometric configuration among the user, APs, transmitter, and the surrounding propagation environment on the performance of Wi-Fi-based HAR.

\subsubsection{Two-Class Gesture Recognition}

\begin{table}[!t]
\centering
\caption{MORIC Binary Hand Motion Recognition Generalization Accuracy (orientation $180^\circ$, ALL APs)}
\resizebox{\columnwidth}{!}{%
\begin{tabular}{c|cccc}
  \hline
  & \textbf{Circle} & \textbf{Left–right} & \textbf{Up–down} & \textbf{Push–pull} \\
  \hline
  \textbf{Circle}      & \cellcolor{gray!50} & $78.8\%\pm11.9\%$ & $91.7\%\pm7.2\%$ & $86.6\%\pm4.7\%$ \\
  \textbf{Left–right}  & \cellcolor{gray!50} & \cellcolor{gray!50} & $88.8\%\pm11.3\%$ & $50.4\%\pm0.9\%$ \\
  \textbf{Up–down}     & \cellcolor{gray!50} & \cellcolor{gray!50} & \cellcolor{gray!50} & $89.2\%\pm9.6\%$ \\
  \textbf{Push–pull}   & \cellcolor{gray!50} & \cellcolor{gray!50} & \cellcolor{gray!50} & \cellcolor{gray!50} \\
  \hline
\end{tabular}
}
\label{table:confusion_matrix}
\vspace{-4mm}
\end{table}

The four gestures in the UTHAMO dataset were deliberately chosen as axis-permuted variants in $3$-D space—e.g., \textit{left–right} versus \textit{up–down}—to evaluate how effectively Wi-Fi-based HAR methods can distinguish between geometrically similar motions. This evaluation is particularly challenging because, unlike camera systems where a reference point such as $(0,0,0)$ can be clearly defined in the environment, Wi-Fi APs lack a fixed spatial reference point in practice. To further investigate this, a binary hand motion classification experiment is conducted to assess which pairs of gestures are easier or more difficult for MORIC to discriminate when the other two gesture classes are excluded.

\begin{figure}[!b]
        \centering		\includegraphics[width=0.88\linewidth]{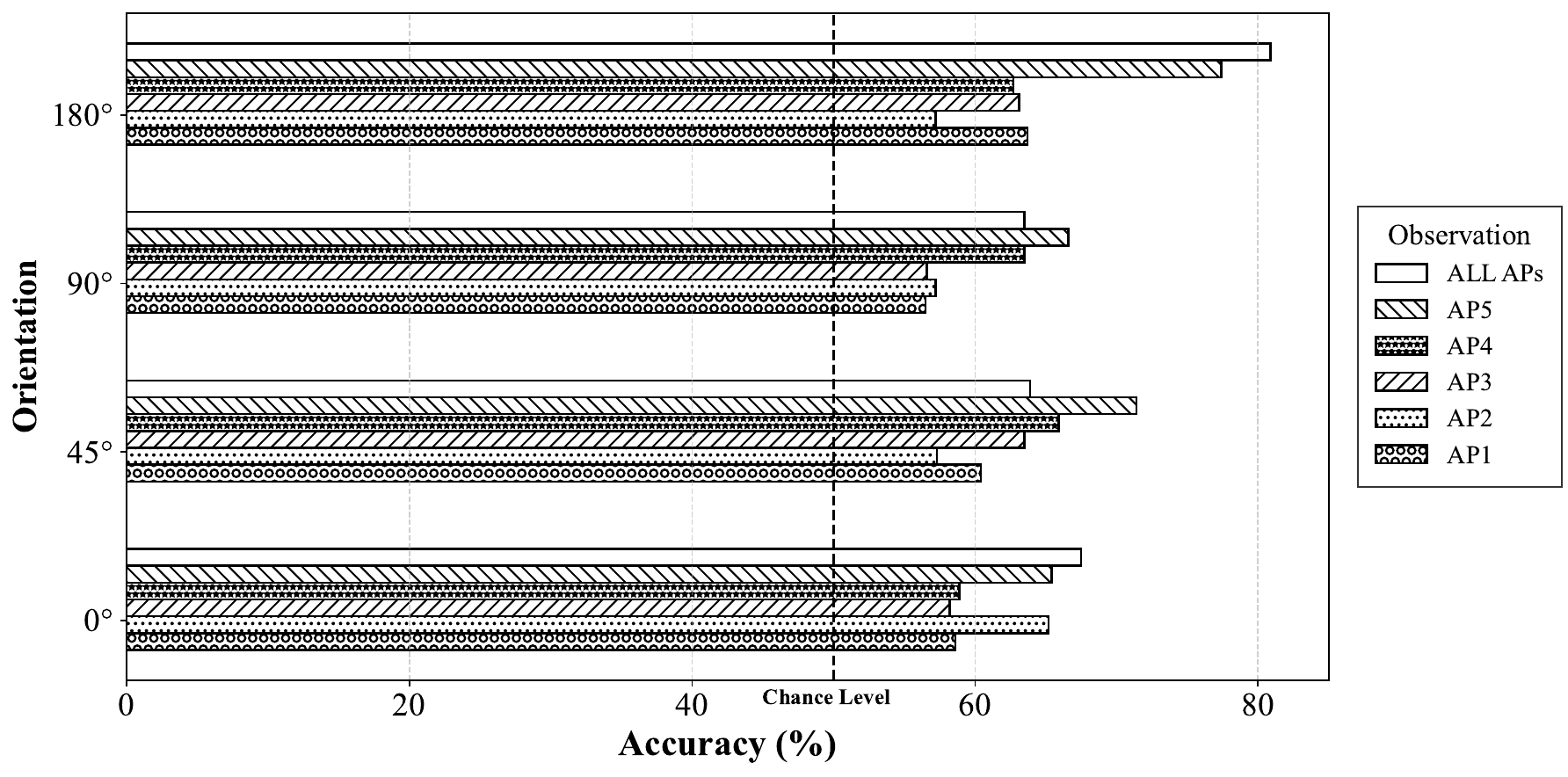}
	
	\caption{Generalization accuracy of MORIC to unseen users averaged over six participants for \textbf{binary hand motion pairs} selected from the six combinations of \textit{circle}, \textit{left–right}, \textit{up–down}, and \textit{push–pull}—across different user orientations.}
	\vspace{-0mm}
\label{figure:binaryhist}
\end{figure}

Fig.~\ref{figure:binaryhist} shows that at orientation $180^\circ$, MORIC achieves a high average generalization accuracy of $80.9\%$ on completely unseen users when using the CSI data from all APs as input to the model.
Consistent with the results from the four-gesture classification, AP~$5$ provides the most motion-related information among the single APs, achieving generalization accuracies ranging from $65.4\%$ at orientation $0^\circ$ to $77.4\%$ at $180^\circ$. For orientation $180^\circ$, Table~\ref{table:confusion_matrix} shows that among the six binary pairs formed from the four gestures, MORIC distinguishes the pairs \textit{up–down vs. circle}, \textit{push–pull vs. up–down}, and \textit{left–right vs. up–down} with high generalization accuracies of $91.7\%$, $89.2\%$, and $88.8\%$, respectively. In contrast, the pair \textit{push–pull vs. left–right} appears to be particularly challenging for the model, as these gestures can look quite similar when the coordinate system is rotated. The gestures \textit{up–down} and \textit{left–right} are more distinguishable, as \textit{left–right} typically involves slight vertical movements at the beginning and end of the gesture, whereas \textit{up–down} consists solely of vertical hand motion. Furthermore, the pair \textit{circle vs. left–right}, with an average accuracy of $78.8\%$, also presents a challenge, probably because people often draw horizontal ellipses in practice rather than perfectly symmetric circles.

\subsubsection{Performance Analysis of Model Calibration}

Although MORIC significantly improves generalization accuracy compared to previous state-of-the-art methods and demonstrates greater robustness to user variability, its performance may still be insufficient for recognizing more than two activities, particularly when the activities are inherently challenging, for practical deployment in real-world applications. Indeed, differences in gesture duration, motion patterns, hand displacement magnitude, and motion speed introduce substantial intra-class variability, which poses a challenge for reliable classification.
\begin{figure}[!t]
    \centering
    \subfloat[\normalfont $4$‐classes\label{fig:cal:a}]{%
        \includegraphics[width=0.48\linewidth]{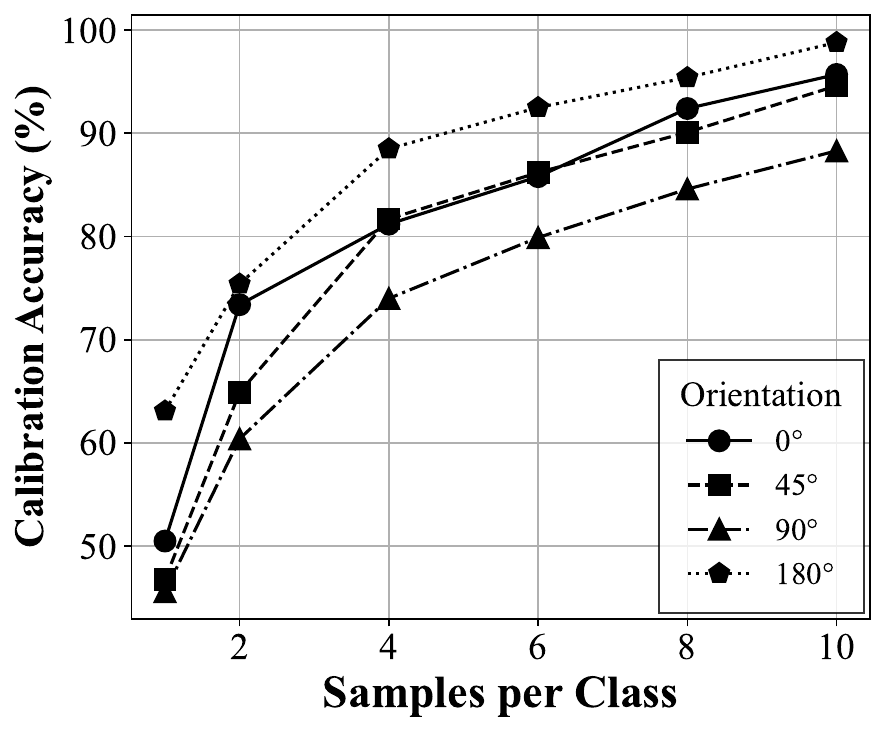}
    }
    \hfill
    \subfloat[\normalfont $2$-classes\label{fig:cal:b}]{%
        \includegraphics[width=0.48\linewidth]{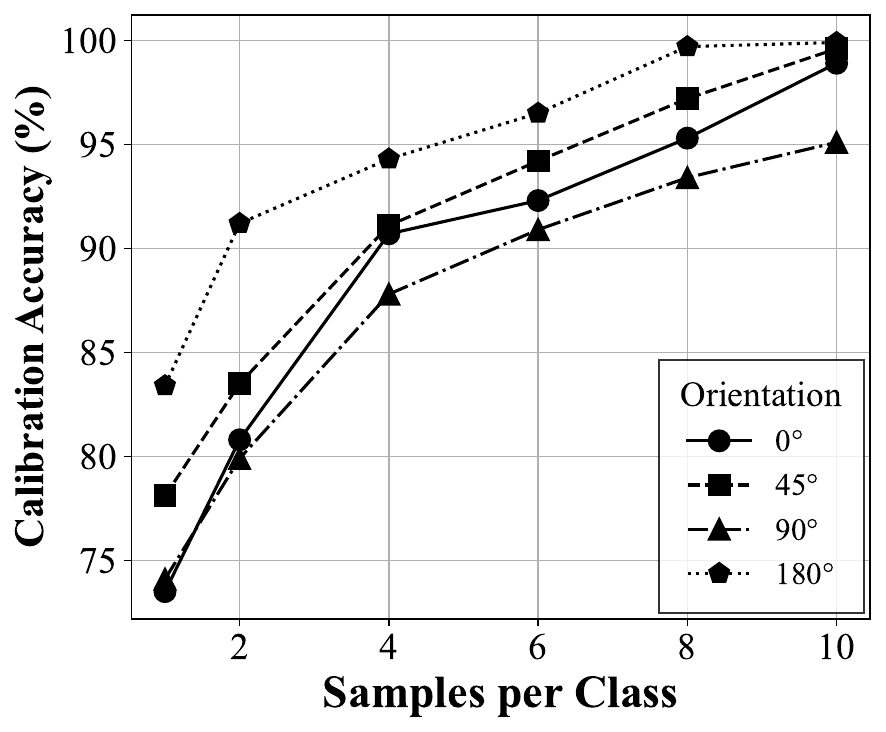}
    }
    \caption{Accuracy of MORIC after calibration with a few samples when AP~$5$ is used as input.}
    \label{figure:calibration}
\end{figure}

Fig.~\ref{figure:calibration} shows that the proposed calibration procedure significantly improves the average accuracy of HAR in both the $4$-class and binary gesture classification scenarios. For the orientation of $180^\circ$, when the model is calibrated using four samples per gesture class, the recognition accuracy for the $4$ classes increases to $88.5\%$ using only SAP~$5$ as input. Increasing the number of calibration samples per class to six further boosts the accuracy to $92.5\%$, and using ten samples per class yields a high accuracy of $98.8\%$. In the binary gesture setting, an accuracy of $94.3\%$ is achieved with four calibration samples per class. These results indicate that the model can accurately recognize small-scale fine-grained hand gestures with as few as four calibration samples per class, highlighting its potential for practical deployment with limited user-specific calibration.

\subsubsection{\texorpdfstring{\revised{Computational Complexity}}{Computational Complexity}}

\revised{To assess the computational cost of MORIC, Table~\ref{table:complexity} reports the average inference-time breakdown for a CSI sample of one AP with three antennas and 64 subcarriers. The evaluation is conducted on a Raspberry~Pi~5 with 8~GB RAM, representing a low-end edge device, and a PC with an Intel Core i9-13900HK processor and 64~GB RAM. The main bottleneck is PSD-based Doppler estimation, which requires approximately \(6.9\)~s with a single thread on the Raspberry~Pi~5; however, because this stage can be parallelized across delay components, using two threads reduces it to approximately \(3.7\)~s. On the Intel Core i9-13900HK PC, the same stage is substantially faster and can further benefit from four-thread execution. The remaining components, including CSI phase compensation, delay-domain decomposition, random-kernel feature extraction, and MORIC classifier inference on CPU, are comparatively lightweight. These results show that the proposed method can run on low- and high-end devices within seconds, and even in less than one second on high-end hardware with parallel execution, making the approach practical for HAR deployment depending on implementation and hardware specifications.}

\begin{table}[!t]
\centering
\caption{\revised{Computational complexity analysis for low- and high-end devices, averaged over 10 runs.}}
\label{table:complexity}
\def\arraystretch{1.05}%
\begin{adjustbox}{width=\linewidth}
\begin{tabular}{lccc}
\hline
\textbf{\revised{Component}} & \textbf{\revised{Threads}} & \textbf{\revised{\makecell{Raspberry~Pi~5\\8~GB RAM}}} & \textbf{\revised{\makecell{PC, Intel Core\\i9-13900HK\\64~GB RAM}}} \\
\hline
\revised{CSI phase compensation} & \revised{--} & \revised{\(\sim 0.10\) s} & \revised{\(\sim 0.01\) s} \\
\revised{Delay-domain decomposition} & \revised{--} & \revised{\(\sim 0.04\) s} & \revised{\(\sim 0.01\) s} \\
\multirow{3}{*}{\revised{\makecell[l]{PSD-based Doppler\\Estimation}}} & \revised{single} & \revised{\(\sim 6.90\) s} & \revised{\(\sim 2.70\) s} \\
 & \revised{two} & \revised{\(\sim 3.70\) s} & \revised{\(\sim 1.20\) s} \\
 & \revised{four} & \revised{--} & \revised{\(\sim 0.85\) s} \\
\revised{Multipath feature extraction} & \revised{--} & \revised{\(\sim 0.28\) s} & \revised{\(\sim 0.11\) s} \\
\revised{MORIC Classifier (on CPU)} & \revised{--} & \revised{\(\sim 0.40\) s} & \revised{\(\sim 0.16\) s} \\
\hline
\end{tabular}
\end{adjustbox}
\end{table}

\subsubsection{Ablation Study}
Table~\ref{table:ablation_moric} presents an ablation study that evaluates the contribution of each component of the MORIC pipeline to the overall activity-recognition performance. The complete MORIC model achieves the highest generalization accuracy of $56.3\%$ with a standard deviation of $9.1\%$, indicating the effectiveness of the complete architecture in extracting motion-relevant features in cross-user settings.
\begin{table}[!b]
\centering
\caption{Ablation study of Generalization Accuracy for MORIC (orientation $180^\circ$, ALL APs)}
\def\arraystretch{1.0}%
\begin{tabular}{lcc}
\hline
\textbf{Configuration}              & \textbf{Average (\%)} & \textbf{SD (\%)} \\
\hline
w/o CSI Phase Compensation                 &  \,27.8\%\,                    &  \,3.4\%\,                 \\
\revised{w/o Delay-Doppler Pre-processing}              &  \,51.4\%\,                    &  \,8.1\%\,                 \\
w/o Max Pooling               &  \,42.5\%\,                    &  \,7.6\%\,                 \\
Phase derivative-based Doppler Est.&  \,49.8\%\,                    &  \,6.5\%\,                 \\
Simple Ridge Classifier  &  39.7\%                    &  \, 8.9\%\,                 \\
\revised{Simple MLP Classifier}  &  \revised{41.2\%}                    &  \revised{8.7\%}                 \\
\revised{End-to-end LSTM Classifier}  &  \revised{37.9\%}                    &  \revised{6.8\%}                 \\
\revised{End-to-end CNN Classifier}  &  \revised{43.8\%}                    &  \revised{8.1\%}                 \\
\hline
\textbf{Full MORIC}                          &  \,\textbf{56.3\%}\,                    &  \,\textbf{9.1\%}\,                 \\
\hline
\end{tabular}
\label{table:ablation_moric}
\end{table}
Removing the CSI phase compensation step results in a substantial performance drop to $27.8\%$, highlighting the importance of mitigating phase distortions caused by STO and SFO. \revised{Similarly, omitting the delay-Doppler pre-processing stage reduces the accuracy to $51.4\%$, confirming the critical role of Doppler domain filtering in suppressing noisy spikes and eliminating low-SNR Doppler velocity representations before MORIC classification.} Excluding the max pooling operation also degrades the performance to $42.5\%$, suggesting that the invariance of MORIC to the input order and repetition is essential for robust HAR. Furthermore, substituting PSD-based Doppler velocity estimation with a simpler phase derivative-based method results in a lower accuracy of $49.8\%$, emphasizing the advantage of a more robust Doppler velocity estimation. Finally, replacing the full classifier with a simple Ridge classifier yields an accuracy of $39.7\%$, demonstrating that the fully designed model architecture is necessary to achieve optimal generalization performance. \revised{Additional classifier ablations further evaluate whether the MORIC feature-extraction and classification design can be replaced by simpler alternatives. The Simple MLP Classifier uses features extracted by random convolutional kernels as input to a two-layer MLP with a hidden-layer size of \(256\), achieving an average accuracy of \(41.2\%\) with an SD of \(8.7\%\). The performance gap between this baseline and MORIC highlights the importance of treating \(\mathcal{F}_P\) as an unordered set of delay-bin representations rather than as a fixed-order concatenated feature vector. Applying a single MLP to the concatenated representations imposes an artificial ordering on the delay-bin features, making the classifier sensitive to arbitrary permutations of \(\mathcal{F}_P\). MORIC avoids this limitation by applying shared MLP heads to all delay-bin representations, followed by max-pooling across the resulting representation-level features. This allows the classifier to retain the strongest evidence for each learned feature regardless of the delay index at which it appears.

The End-to-end LSTM Classifier directly receives the Doppler projections as time-series inputs and performs both feature extraction and classification using a two-layer LSTM with a hidden size of \(256\), achieving \(37.9\%\pm6.8\%\). The End-to-end CNN Classifier treats the Doppler projections as an image, where one axis corresponds to delay and the other to time, and applies a 2-D CNN for both feature extraction and classification. This CNN uses three convolutional blocks with \(3\times3\) kernels and channel dimensions \(1\!\rightarrow\!32\!\rightarrow\!64\!\rightarrow\!128\), achieving \(43.8\%\pm8.1\%\). These results show that all three simpler alternatives underperform MORIC, highlighting the importance of the proposed pipeline for Wi-Fi-based HAR, which combines temporal feature extraction using random convolutional kernels with order-invariant classification of Doppler velocity projections.}

\section{\texorpdfstring{\revised{Limitations and Future Work}}{Limitations and Future Work}}\label{section:limitations}
\revised{The proposed MORIC framework significantly improves cross-user generalization accuracy compared with previous Wi-Fi-based HAR methods, indicating strong potential for future deployment in practical and robust Wi-Fi sensing systems. However, the present study still has several limitations that should be addressed in future research.}

\revised{First, MORIC is designed to reduce sensitivity to static environmental settings and emphasize motion-induced signal variations. Nevertheless, the results show that the relative placement of the APs with respect to the user and transmitter can still affect HAR performance. Due to the limited bandwidth of sub-\(7\) GHz Wi-Fi, each delay bin may aggregate multiple physical paths with similar propagation delays but different interaction directions. The AP location can therefore influence how concentrated these paths are within each delay bin, which corresponds to different effective values of the vMF concentration parameter \(\kappa\). When the contributing paths are concentrated around a dominant direction, the first-order path-length approximation in~\eqref{eq:linear3d} and the phase-to-velocity relation in~\eqref{eq:phase_laplace} support a sharper interpretation of the extracted Doppler velocity as an approximately linear projection of the motion velocity. However, when the paths are more dispersed or multimodal, this linear projection interpretation becomes weaker, and the extracted Doppler velocity represents a mixed and blurred superposition of motion projections from several directions. This indicates that, although MORIC improves robustness to static background effects, practical deployment is still influenced by sensing geometry and the limited bandwidth of Wi-Fi, which constrains delay resolution and affects the separability of multipath components.
}

\revised{In addition, while the UTHAMO dataset was collected in a practical office environment, the experimental protocol still included controlled aspects, including fixed device placement, scripted gestures, fixed trial timing, and a static background without uncontrolled human movement. Thus, the reported results should be interpreted as evidence of improved cross-user generalization in a fixed indoor deployment, rather than as a complete evaluation of fully uncontrolled real-world operation.}

\revised{Second, the theoretical formulation assumes that the environmental reflections and their corresponding observation directions, modeled through the vMF distribution, remain approximately fixed during the seconds-long activity window. This assumption is necessary because explicitly modeling random and unpredictable time-varying scattering caused by arbitrary human motion and environmental perturbations is not tractable within the current framework. In real-world scenarios, however, the reflections themselves may also change as motion occurs in the environment, which can alter the effective multipath observations during the same activity.}

\revised{Third, the current theoretical formulation is based on random multipath propagation in an enclosed environment and does not assume a fixed reference coordinate point in 3-D space. This makes the method well matched to commodity Wi-Fi systems, where APs generally do not provide a camera-like global coordinate frame. However, the absence of such a fixed spatial reference also makes it challenging to distinguish between gestures that become nearly equivalent under rotations of the \(x\), \(y\), and \(z\) axes. For example, \textit{left--right} and \textit{push--pull} motions can produce highly similar observations when the effective sensing axes are rotated, which limits separability for very similar gesture pairs.}

\revised{Fourth, although the proposed method produces a substantial improvement in recognition accuracy, an optional calibration stage with only a few user-specific samples may still be required when the system is intended to operate reliably in real-world HAR settings. The calibration results show that a small number of samples can greatly improve performance, but this also indicates that fully calibration-free robust HAR remains an open challenge. Addressing these limitations in future work can further pave the way toward completely practical and robust Wi-Fi sensing.}

\section{Conclusion}\label{section:conclusion}
\revised{This study introduces MORIC, a robust framework for Wi-Fi-based human activity and gesture recognition that significantly improves cross-user generalization accuracy. By transforming CSI into the delay domain and extracting Doppler velocity projections, the proposed method establishes a new perspective on Wi-Fi sensing by capturing motion-induced dynamics while suppressing the influence of static environmental components. These projections characterize human motion from multiple random observation directions, forming a discriminative and robust representation space. To effectively exploit this representation while accounting for multipath randomness, MORIC is designed to be invariant to the stochastic nature of indoor multipath propagation.}

Empirical evaluations on a challenging hand gesture dataset collected in this study confirm the effectiveness of the proposed approach for cross-user generalization. MORIC consistently outperforms state-of-the-art methods in both single- and multi-AP settings, achieving higher recognition accuracy under domain shifts, even with a single AP. Incorporating a small number of calibration samples further improves performance, highlighting the potential of the proposed method to bring practical real-world Wi-Fi-based HAR deployment closer to realization.

\section*{\revised{AI Use Disclosure}}
\revised{Generative AI tools were used to assist with language editing, response drafting, and improving the clarity of technical descriptions in the manuscript. The authors reviewed, edited, and verified the resulting content and take full responsibility for the final manuscript.}

\bibliographystyle{IEEEtran}

\bibliography{refs}



\newpage

\begin{IEEEbiography}[{\includegraphics[width=1in,height=1.25in,clip,keepaspectratio]{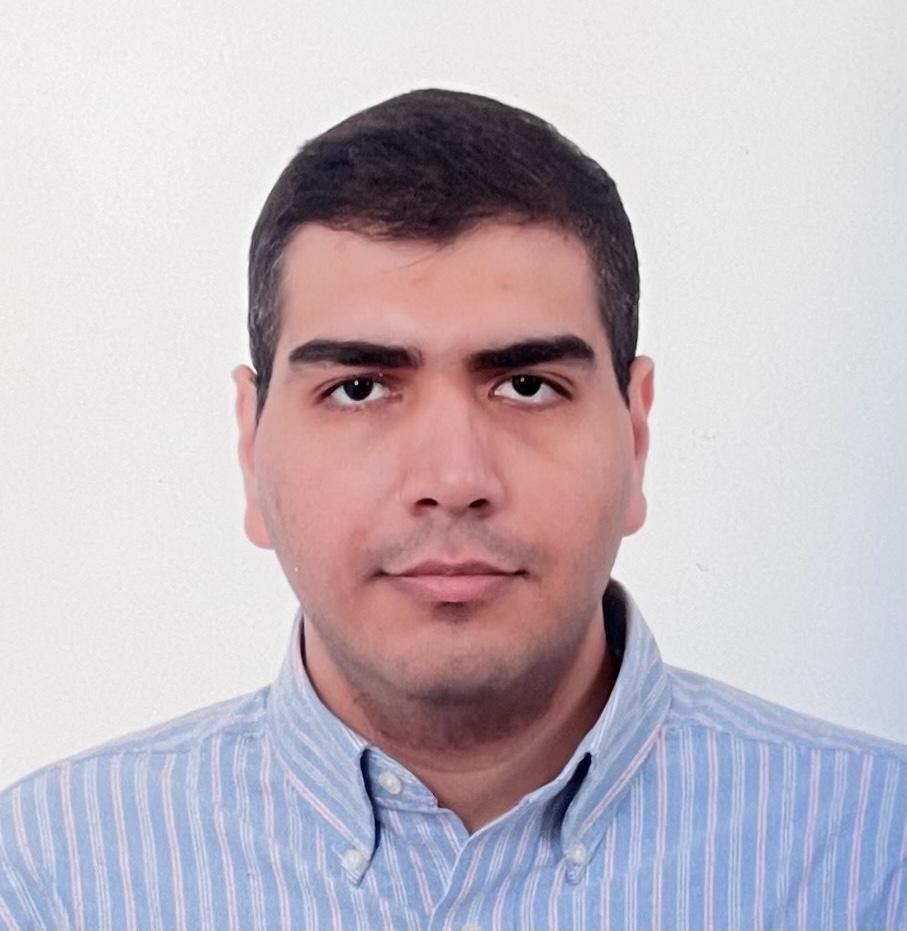}}]{Navid Hasanzadeh}
\revised{is currently a Ph.D. candidate in the Edward S. Rogers Sr. Department of Electrical and Computer Engineering at the University of Toronto, Toronto, ON, Canada. 

His research interests include wireless sensing, machine learning for healthcare, and interpretable and explainable AI. His work integrates signal processing and machine learning to develop robust methods for Wi-Fi-based human activity and gesture recognition, digital and wearable health monitoring, and interpretable AI, with a focus on improving generalization, transparency, and reliability across diverse real-world settings.}
\end{IEEEbiography}

\begin{IEEEbiography}[{\includegraphics[width=0.9in,height=1.25in,clip,keepaspectratio]{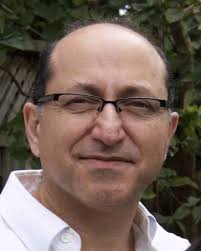}}]{Shahrokh Valaee} \revised{is a Professor with the Edward S. Rogers Sr. Department of Electrical and Computer Engineering, University of Toronto, and the holder of the Nortel Chair of Network Architectures and Services. He is the Founder and the Director of the Wireless Innovation Research Laboratory (WIRLab) at the University of Toronto. 

Professor Valaee was the TPC Co-Chair and the Local Organization Chair of the IEEE Personal Mobile Indoor Radio Communication (PIMRC) Symposium 2011, the TPC Co-Chair of ICT 2015, PIMRC 2017, and the Track Co-Chair of WCNC 2014, PIMRC 2020, and VTC Fall 2020. He was the co-chair of the organizing committee for PIMRC 2023. He is currently a member of the Steering Committee of IEEE PIMRC. From December 2010 to December 2012, he was the Associate Editor of the IEEE Signal Processing Letters. From 2010 to 2015, he served as an Editor of IEEE Transactions on Wireless Communications. Currently, he is an Editor of IEEE Transactions on Wireless Communications and an Associate Editor of the Journal of Computer and System Sciences. From 2021 to 2023, he was a Distinguished Lecturer of the IEEE Communications Society. Currently, he serves as a Distinguished Lecturer for the IEEE Vehicular Technology Society. He was the co-recipient of the best paper award in the IEEE Machine Learning for Signal Processing (MLSP) 2020 workshop. Professor Valaee is a Fellow of the Engineering Institute of Canada and a Fellow of IEEE. }

\end{IEEEbiography}

\vfill

\newpage
\revised{\phantomsection
\section*{Appendix A\\
PROOF OF PROPOSITION~1}
\label{Appendix:A}
\addcontentsline{toc}{section}{Appendix A: PROOF OF PROPOSITION~1}

\begin{proof}
The updated position of the moving point is $P_{t} = P_0 + \mathbf{v}\,t$. The distance from $O$ to the updated position is
\[
\|P_{t} - O\| = \|P_0 + \mathbf{v}t - O\| = \|\mathbf{r}_O + \mathbf{v}t\|.
\]
Using the squared norm identity,
\[
\|\mathbf{r}_O + \mathbf{v}t\|^2 
= \|\mathbf{r}_O\|^2 
+ 2 \mathbf{r}_O \cdot \mathbf{v}  t 
+ \|\mathbf{v}\|^2 t^2,
\]
so the total change is
\[
\Delta L_O = 
\sqrt{
d_O^2 + 2 \mathbf{r}_O \cdot \mathbf{v} t + \|\mathbf{v}\|^2 t^2
} - d_O.
\]

For small $\|\mathbf{v}\|\, t$, apply the Taylor expansion
\[
\sqrt{d_O^2 + a} = d_O + \frac{a}{2d_O} + \mathcal{O}(a^2),
\]
with \( a = 2 \mathbf{r}_O\cdot \mathbf{v} t + \|\mathbf{v}\|^2 t^2 \). Discarding the second-order term gives
\[
\Delta L_O \approx \frac{ \mathbf{v} \cdot \mathbf{r}_O}{d_O}\,t = \|\mathbf{v}\|\, t \cos\theta_O. 
\]

\end{proof}}

\revised{\phantomsection
\section*{Appendix B\\
PROOF OF PROPOSITION~3}
\label{Appendix:B}
\addcontentsline{toc}{section}{Appendix B: PROOF OF PROPOSITION~3}

\begin{proof}
The map $\mathbf{r}\mapsto\mathbf{v}(s)\,\!\cdot\!\,\mathbf{r}$ is linear and therefore smooth.  
The exponential $\exp\,\!\bigl(j\alpha\,\mathbf{v}(s)\!\cdot\!\mathbf{r}\bigr)$, with $\alpha = 2\pi\,t\, f_c/c$, is a smooth composition of analytic functions.
Since $\beta_i$ is smooth and the product of smooth functions remains smooth, $g(\mathbf{r})$ belongs to $C^\infty(\mathbb{S}^2)$.

\end{proof}}

\revised{\phantomsection
\section*{Appendix C\\
PROOF OF PROPOSITION~4}
\label{Appendix:C}
\addcontentsline{toc}{section}{Appendix C: PROOF OF PROPOSITION~4}

\begin{proof}
Using local coordinates around \(\mathbf{m}\), write \( \mathbf{r} = \mathbf{m} + \delta \mathbf{r} \), with \( |\delta \mathbf{r}| = \mathcal{O}(\kappa^{-1/2}) \). Expanding \( f(\mathbf{r}) \) using Taylor's theorem yields
\begin{equation*}
f(\mathbf{r}) = f(\mathbf{m}) + \nabla f|_{\mathbf{m}} \, \delta \mathbf{r} + \mathcal{O}(|\delta \mathbf{r}|^2).
\end{equation*}
Due to the symmetry of \( p_{\kappa}(\mathbf{r}) \) around \(\mathbf{m}\), the linear term vanishes under integration. Moreover, since \( |\delta \mathbf{r}| = \mathcal{O}(\kappa^{-1/2}) \), the quadratic term satisfies \( |\delta \mathbf{r}|^2 = \mathcal{O}(1/\kappa) \). Therefore, the second-order term contributes \( \mathcal{O}(1/\kappa) \), and
\begin{equation*}
\iint_{\mathbb{S}^2} f(\mathbf{r}) p_{\kappa}(\mathbf{r}) \, d\mathbf{r} = f(\mathbf{m}) + \mathcal{O}\left(\frac{1}{\kappa}\right).
\end{equation*}

\end{proof}}

\end{document}